\documentclass[a4paper,aps,prd,eqsecnum,amssymb,nofootinbib,floatfix,twocolumn,showpacs]{revtex4-1}

\usepackage{amsmath}
\usepackage{dcolumn}
\usepackage{graphicx}
\usepackage{multirow}

\allowdisplaybreaks

\newcommand{\bdm}{\begin{displaymath}}
\newcommand{\edm}{\end{displaymath}}
\newcommand{\be}{\begin{equation}}
\newcommand{\ee}{\end{equation}}
\newcommand{\bse}{\begin{subequations}}
\newcommand{\ese}{\end{subequations}}
\newcommand{\ba}{\begin{array}}
\newcommand{\ea}{\end{array}}

\newcommand{\btau}{\boldsymbol{\tau}}
\newcommand{\btheta}{\boldsymbol{\theta}}
\newcommand{\bxi}{\boldsymbol{\xi}}
\newcommand{\bzeta}{\boldsymbol{\zeta}}

\newcommand{\Ca}{C_\textrm{a}}
\newcommand{\Ce}{C_\textrm{e}}

\newcommand{\chii}{\chi_\text{i}}
\newcommand{\cmin}{C_\textrm{min}}
\newcommand{\cmine}{C_\textrm{min}^\textrm{e}}
\newcommand{\cmini}{C_\textrm{min}^\textrm{i}}

\newcommand{\F}{\mathcal{F}}
\newcommand{\Fc}{\text{F}_\text{c}}
\newcommand{\Fs}{\text{F}_\text{s}}
\newcommand{\fc}{f_\text{c}}
\newcommand{\fN}{f_\text{N}}

\newcommand{\md}{{\mathrm{d}}}

\newcommand{\mi}{{\mathrm{i}}}

\newcommand{\To}{T_\text{o}}

\newcommand{\ti}{t_\text{i}}

\begin{document}

\title{Banks of templates
for directed searches of gravitational waves
from~spinning~nuetron~stars}

\author{Andrzej Pisarski}
\email{andrzej@alpha.uwb.edu.pl}
\affiliation{Faculty of Physics,
University of Bia{\l}ystok,
Lipowa 41, 15--424 Bia{\l}ystok, Poland}

\author{Piotr Jaranowski}
\email{pio@alpha.uwb.edu.pl}
\affiliation{Faculty of Physics,
University of Bia{\l}ystok,
Lipowa 41, 15--424 Bia{\l}ystok, Poland}

\author{Maciej Pietka}
\email{map@alpha.uwb.edu.pl}
\affiliation{Faculty of Physics,
University of Bia{\l}ystok,
Lipowa 41, 15--424 Bia{\l}ystok, Poland}

\date{\today}

\begin{abstract}

We construct efficient banks of templates suitable
for directed searches of almost monochromatic gravitational waves
originating from spinning nuetron stars in our Galaxy
in data being collected by currently operating interferometric detectors.
We thus assume that the position of the gravitational-wave source in the sky is known,
but we do not assume that the wave's frequency and its derivatives are a priori known.
In the construction we employ simplified model of the signal with constant amplitude
and phase which is a polynomial function of time.
All our template banks enable usage of the fast Fourier transform algorithm
in the computation of the maximum-likelihood $\F$-statistic for nodes of the grids defining the bank.
We study and employ the dependence of the grid's construction
on the choice of the position of the observational interval
with respect to the origin of time axis.
We also study the usage of the fast Fourier transform algorithms
with non-standard frequency resolutions achieved by zero padding or folding the data.
In the case of the gravitational-wave signal with one spindown parameter included
we have found grids with covering thicknesses which are only 0.1\%--16\% larger
than the thickness of the optimal two-dimensional hexagonal covering.

\end{abstract}


\maketitle

\section{Introduction and summary}

Rotating neutron stars in our Galaxy are expected sources
of almost monochromatic gravitational waves
which are looked for in the data being collected
by currently operating ground-based interferometric detectors
LIGO \cite{LIGO}, Virgo \cite{Virgo}, GEO600 \cite{GEO}, and TAMA \cite{TAMA}.
In the present paper we consider the problem of construction
of efficient banks of templates needed to detect
gravitational-wave signals originating from spinning neutron stars.
We assume that the problem of detection of the signal and of estimation of its parameters
is based on the \textit{maximum-likelihood} (ML) principle
and we also assume that the noise in the detector is Gaussian and stationary
(the detailed exposition of the ML detection in Gaussian noise can be found
e.g.\ in Chapter 6 of the monograph \cite{JKbook},
see also \cite{W71} and the review article \cite{JKlrr}).
Data analysis tools and algorithms needed to perform,
within the ML approach, an \textit{all-sky} search
for almost monochromatic gravitational-wave signals
(i.e.\ the search which assumes
that the position of the source in the sky is not known)
were developed in detail in the series
of papers \cite{JKS98,JK99,JK00,ABJK02,ABJPK10}
(see also Refs.\ \cite{BCCS98,BC99}).
In our paper we restrict ourselves to \emph{directed} searches
for almost monochromatic signals,
i.e.\ we assume that the position of the source in the sky is known.
We do not assume however that the frequency and the spindown parameters
of the gravitational-wave signal are known.
Directed searches should thus be distinguished from \emph{targeted} searches,
in which it is additionally assumed that the frequency and the spindown parameters
of the gravitational wave are also known (see Refs.\ \cite{PK09,JK10}
for discussions of statistics by means of which one can test
whether data contains such signal).
Several targeted searches were already performed with data collected
by the LIGO and GEO600 detectors \cite{LSC04,LSC05,LSC07,LSC08,LSC10}.

In the ML approach one considers the likelihood ratio $\Lambda[x;\btheta]$
which is a function of the data $x$
and the parameters $\btheta$ of the gravitational-wave signal we are looking for.
Detection of the signal relies on the computation of $\Lambda[x;\btheta]$
maximized over all possible values of the parameters $\btheta$
and comparing this maximum with a threshold.
In the case of directed searches the unknown parameters $\btheta$
can be divided into two groups, $\btheta=(\mathbf{A},\bxi)$.
The first group $\mathbf{A}$ consists of four \emph{extrinsic} or \emph{amplitude} parameters:
an overall amplitude of the waveform, its initial phase, the polarization angle of the wave,
and the inclination angle of the star's rotation axis with respect to the line of sight.
The second group $\bxi$ contains \emph{intrinsic} or \emph{phase} parameters:
the frequency of the wave and the spindown parameters.
Maximization of the $\Lambda$ with respect to amplitude parameters $\mathbf{A}$ can be done analytically
(by solving the set of ML equations $\partial\Lambda/\partial\mathbf{A}=0$
with respect to $\mathbf{A}$)
and the \textit{$\F$-statistic} is defined as the logarithm of the likelihood ratio $\Lambda$
after replacing in $\Lambda$ the amplitude parameters $\mathbf{A}$
by their ML estimators $\hat{\mathbf{A}}$:
$\F[x;\bxi]:=\ln\Lambda[x;\hat{\mathbf{A}},\bxi]$.
Maximization of the $\F$-statistic over the phase parameters $\bxi$ can be done only numerically.
To find this maximum one needs to construct a \textit{bank of templates} in the space of the parameters $\bxi$
on which the $\F$-statistic depends.
The bank of templates is defined by a discrete set of points, i.e.\ a \emph{grid} in the parameter space
chosen in such a way, that for any possible signal there exists a grid point
such that the expectation value of the $\F$-statistic (which is a random variable
as it depends on the detector's noise which is a stochastic process) computed for the parameters
of this grid point is not less than a certain fixed minimal value
(assuming that the minimal value of the signal-to-nise ratio is also a priori fixed).

In the series of papers \cite{JKS98,JK99,JK00,ABJK02,ABJPK10}
it was argued and checked by numerical simulations
that in the case of an all-sky search in the construction of the bank of templates
one can employ a simplified model of the gravitational-wave signal, the so called
\textit{linear phase model}, in which the amplitude of the signal is assumed to be constant and
the signal's phase is a linear function of the unknown parameters (this model
was introduced in Sec.\ V B of Ref.\ \cite{JK99}). In Sec.\ V E of \cite{JK99}
it was checked that the linear model reproduces well the covariance matrix
(defined as the inverse of the Fisher matrix)
for the ML estimators of the signal's parameters of the exact gravitational-wave signal.
Also in Ref.\ \cite{JK99} (Sec.\ V D and Appendix C there) the \textit{polynomial phase model} was
introduced in which the signal's amplitude is constant and the phase is a polynomial function
of time. It was found (in  Sec.\ V E of \cite{JK99}) that the polynomial phase model reproduces
very well the covariance matrix of the signal's parameters of the exact model in the case of directed searches.
This indicates that the polynomial model can be used in the construction of the banks of templates
for directed searches and this model is accepted in the present paper.

In the current paper we assume that the observational interval is of the form
$\langle\ti-\To/2;\,\ti+\To/2\rangle$ (where $\To$ is the length of observation time),
and we study the dependence of the construction of banks of templates on the choice
of the parameter $\ti$ (or its dimensionless version $\chii:=\ti/\To$)
which fixes the position of the observational interval with respect to the origin of time axis.

The organization and the main results achieved in the paper are as follows.
In Sec.\ II we introduce the polynomial phase model of the gravitational-wave signal.
We consider here the phase with only one spindown parameter included.
For this model we compute the $\F$-statistic and its expectation value
in the case when the data contains the gravitational-wave signal.
This expectation value depends [see the crucial Eq.\ \eqref{E1Fc} below] on the signal-to-noise ratio $\rho$
and on the value of the \textit{autocovariance function} $C_0(\bxi,\bxi')$ of the $\F$-statistic
(the subscript `0' indicates that the autocovariance is calculated in the case when data is a pure noise)
computed for the intrinsic parameters of the template ($\bxi$)
and the gravitational-wave signal ($\bxi'$), respectively.
The signal-to-noise ratio $\rho$ we can not control,
therefore to construct bank of templates
one needs to choose some minimum value $\cmin$ of the autocovariance function $C_0$
and look for such a grid of points that for any point $\bxi'$ in the intrinsic parameter space
there exists a grid node $\bxi$ such that the autocovariance $C_0(\bxi,\bxi')$ computed for the parameters
$\bxi$ and $\bxi'$ is not less than $\cmin$.
The autocovariance $C_0(\btau)$ (for polynomial phase model
it depends on $\bxi$, $\bxi'$ only through the difference $\btau:=\bxi-\bxi'$)
can be expressed in terms of Fresnel integrals
or one can use an approximate formula for $C_0$ by taking the Taylor
expansion (up to the second-order terms) of $C_0$ around its maximum at $\btau=\mathbf{0}$.
In Fig.\ \ref{isoheights} and Table \ref{table1}
we compare these two ways of computing the autocovariance.

In Sec.\ III we consider banks of templates necessary
to perform detection of almost monochromatic gravitational-wave signals
with polynomial phase. We first formulate the problem of constructing bank of templates
as a problem of finding optimal covering of the signal's parameter space by means of
identical ellipses (defined as isoheights of the autocovariance function $C_0$ of the $\F$-statistic)
and introduce some mathematical notions related with coverings.
We are interested in such searches for almost periodic gravitational-wave signals
for which the number of grid points in the parameter space is very large
and the time needed to compute the $\F$-statistic for all grid nodes is long.
Then it is crucial to use in the computation the fast numerical algorithms.
Because the computation of the $\F$-statistic involves calculation of the Fourier transform,
one would like to use the \emph{fast Fourier transform} (FFT) algorithm.
As is known, the FFT algorithm computes the values of the \emph{discrete Fourier transform} (DFT)
of a time series for a certain set of discrete frequencies called the \emph{Fourier frequencies}.
Thus it will be possible to use the FFT algorithm in computation of the $\F$-statistic,
provided the grid points will be arranged in such a way,
that the frequency coordinates of these points
will all coincide with the Fourier frequencies.
We have constructed two different families of grids which fulfill this requirement.
Our constructions were motivated by a simple observation
(see Fig.\ \ref{isoheights})
that the shape of the autocovariance ellipse
strongly depends on the value of the parameter $\chii$:
the larger $|\chii|$ is, the more alongated (along the frequency axis) this ellipse is.
The usage of alongated ellipses suggests that the $\F$-statistic could be computed
for smaller number of frequency values than the standard DFT algorithm computes.
Therefore we have employed some modifications of the DFT algorithm
leading to non-standard frequency resolutions achieved by zero padding or folding the data
(this is discussed in Appendix A).
Details of the grids construction are presented in Appendix \ref{grids},
where also the components of the basis vectors spanning the grids are given
(in Tables \ref{tableG1}--\ref{tableG2o}).

The construction of efficient banks of templates
for gravitational-wave searches
was recently discussed in Ref.\ \cite{MPP09}
(see also \cite{P07,HAS09,MV10,R10}),
where random template banks and relaxed lattice coverings were considered.
As explained above we are interested in searches involving
data streams so long, that the time performance of the search crucially
depends on the ability of using the FFT algorithm. This enforces
the above-mentioned constraint which is not fulfilled by the grids
considered in Ref.\ \cite{MPP09}. Therefore our work can be considered
as being complementary to the studies performed in Ref.\ \cite{MPP09}.
The grid fulfilling the constraint was constructed in Sec.\ IV of Ref.\ \cite{ABJPK10}
in the case of all-sky search for almost monochromatic gravitational waves.
In the search considered in \cite{ABJPK10} the signal's parameter space is 4-dimensional
and the polynomial phase model can not be employed. However,
in Sec.\ IV C of \cite{ABJPK10}, as an example of application of general
algorithm devised in \cite{ABJPK10} to construct constrained grids,
the two-dimensional grid for searches of signals with polynomial phase is considered.
This grid has thickness equal to $\sim$1.8, whereas grids constructed
in the current paper have thicknesses of $\sim$1.2
(all grids found by us have covering thicknesses which are only 0.1\%--16\% larger
than the thickness of the optimal two-dimensional hexagonal covering).

In Sec.\ IV of the current paper we present the results of numerical simulations we performed
to study the quality of grids constructed in Appendix B.
These results are contained in Table \ref{simula}.
From our simulations it follows that when considering the efficiency of a grid
it is not enough to worry only about the thickness of the covering related with the grid.
We see from Table \ref{simula} that for lattices which have the same thickness of covering,
but one of them has basis vectors which are orthogonal to each other,
the time used to compute the $\F$-statistics is at least two times shorter
than the corresponding time for the lattice with non-orthogonal basis vectors.

\section{Autocovariance function of the  ${\cal F}$-statistic}

We assume that the noise $n$ in the detector is an additive, stationary, Gaussian,
and zero-mean continuous stochastic process.
Then the logarithm of the likelihood function is given by
\be
\label{001}
\ln\,\Lambda[x] = (x|h)-\frac{1}{2}\,(h|h),
\ee
where $x$ denotes the data from the detector,
$h$ is the deterministic signal we are looking for in the data,
and $(\cdot\,|\,\cdot\,)$ is the scalar product between waveforms defined by
\be
\label{001a}
(h_{1}|h_{2}) := 4\,\text{Re}
\int_{0}^{\infty} \frac{\tilde{h}_1(f)\,{\tilde{h}}_2^*(f)}{S_n(f)}\: \md f,
\ee
where \hspace*{1ex}$\tilde{}$\hspace*{1ex} stands for the Fourier transform,
* denotes complex conjugation,
and $S_{n}$ is the \textit{one-sided spectral density}
(defined for frequencies $0\le f<+\infty$) of the detector's noise $n$.

We are interested in \textit{almost monochromatic} signals,
i.e.\ such signals for which the modulus of the Fourier
transform is concentrated (for frequencies $f\ge0$) around some `central' frequency $\fc>0$
and $S_{n}$ is a slowly changing function of $f$ in the vicinity of the frequency $\fc$.
If both waveforms $h_1$ and $h_2$ in Eq.\ \eqref{001a} have Fourier transforms
concentrated around the same frequency $\fc$,
then we can replace $S_{n}(f)$ in the integrand of \eqref{001a} by $S_{n}(\fc)$ and,
after employing the Parseval's theorem, approximate the scalar product by
\be
\label{002}
(h_{1}|h_{2}) \cong \frac{2}{S_{n}(\fc)}\,
\int\limits_{\ti-\To/2}^{\ti+\To/2}h_{1}(t)\,h_{2}(t)\:\md t
= \frac{2\,\To}{S_{n}(\fc)}\langle h_{1}h_{2}\rangle.
\ee
Here $\langle\ti-\To/2;\,\ti+\To/2\rangle$ denotes observational interval,
so $\To$ is the length of observation time,
and $\ti-\To/2$ is the moment at which the observation begins.
The time averaging operator $\langle\,\cdot\,\rangle$ is defined by
\be
\label{003}
\langle h \rangle := \frac{1}{\To}\int\limits_{\ti-\To/2}^{\ti+\To/2}h(t)\,\md t.
\ee
Using the formula \eqref{002} we can write the log likelihood ratio from Eq. \eqref{001} as
\be
\label{004}
\ln\Lambda[x] \cong \frac{2\,\To}{S_{n}(\fc)}\,
\Big(\langle x\,h \rangle -\frac{1}{2}\langle h^{2}
\rangle\Big).
\ee

We restrict to almost monochromatic signals with frequency drift
given by the linear-in-time relation,
\be
\label{004a}
f(t)=f_{0}+\dot{f}_{0}\,t,
\ee
so $f_{0}$ and $\dot{f}_{0}$ is the frequency and the first derivative of the frequency with respect to time,
respectively, both taken at the time $t=0$.
The instantaneous frequency is related to the phase $\Psi$ of the signal by
\be
\label{007}
f(t)=\frac{1}{2\pi}\frac{\md\,\Psi(t)}{\md\,t}.
\ee
Making use of Eqs. \eqref{004a} and \eqref{007}, we obtain the time dependence of the phase $\Psi$:
\be
\label{007a}
\Psi(t) = 2\pi\int_{0}^{t}f(t')\,\md\,t'+\Psi_{0}
= 2\pi\,\left(f_{0}\,t+\frac{1}{2}\,\dot{f}_{0}\,t^2\right)
+ \Psi_{0},
\ee
where $\Psi_{0}$ is the value of the phase $\Psi$ at the time $t=0$.
It is convenient to define the \textit{dimensionless} parameters
\be
\label{007c}
\omega_{0} := 2\pi\,f_{0}\,\To,
\qquad
\omega_{1} := \pi\,\dot{f}_{0}\,\To^{2}.
\ee
Using these parameters we can write the phase $\Psi$ in the form
\be
\label{007d}
\Psi(t;\Psi_{0},\omega_{0},\omega_{1})=\Phi(t;\omega_{0},\omega_{1})+\Psi_{0},
\ee
where
\be
\label{008}
\Phi(t;\omega_{0},\omega_{1})=\omega_{0}\frac{t}{\To}+\omega_{1}\bigg(\frac{t}{\To}\bigg)^2.
\ee
We further assume that the gravitational-wave signal $h$ we are looking for
has a constant amplitude $h_{0}$, so it can be written in the form
\be
\label{005}
h(t;h_{0},\Psi_{0},\omega_{0},\omega_{1})
= h_{0}\,\sin\big(\Phi(t;\omega_{0},\omega_{1})+\Psi_{0}\big).
\ee
Let us collect the parameters of the phase $\Phi$
into a $2$-dimensional vector $\bxi$,
\be
\label{009}
\bxi := (\omega_{0},\omega_{1}).
\ee
Then the gravitational-wave signal $h$, Eq.\ \eqref{005},
we can shortly write as
\be
\label{010}
h(t;h_0,\Psi_0,\bxi)
= h_{0}\,\sin\big(\Phi(t;\bxi)+\Psi_{0}\big).
\ee

It is easy to maximize the likelihood ratio \eqref{004} for the signal \eqref{010} with respect to
the parameters $h_{0}$ and $\Psi_{0}$. To do this it is convenient to introduce the new parameters
$h_{1}$ and $h_{2}$:
\be
\label{012}
h_{1} := h_{0}\cos{\Psi_{0}},
\quad
h_{2} := h_{0}\sin{\Psi_{0}},
\ee
and rewrite the signal $h$ in the form
\be
\label{013}
h(t;h_{1},h_{2},\bxi) = h_{1}\sin\Phi(t;\bxi)+h_{2}\cos\Phi(t;\bxi).
\ee
Making use of Eq.\ \eqref{013} and the identities
$2\sin\alpha\cos\alpha=\sin2\alpha$,
$\cos^{2}\alpha=\frac{1}{2}\,(1+\cos2\alpha)$,
$\sin^{2}\alpha=\frac{1}{2}\,(1-\cos2\alpha)$
we can represent the time average $\langle{h^{2}}\rangle$ as
\be
\label{015}
\langle h^{2} \rangle=\frac{1}{2}(h_{1}^{2}+h_{2}^{2})+h_{1}h_{2}\langle \sin2\Phi \rangle
+ \frac{1}{2}(h_{2}^{2}-h_{1}^{2})\langle\cos2\Phi\rangle.
\ee
For observation times longer than few hours
and for almost monochromatic signals
with frequency of the order of hundreds or thousands hertz,
we can make approximation
\be
\label{016}
\langle\sin2\Phi\rangle \cong 0,
\quad
\langle\cos2\Phi\rangle \cong 0.
\ee
Then Eq.\ \eqref{015} can be approximated by
\be
\label{017}
\langle h^{2} \rangle\cong\frac{1}{2}\,(h_{1}^{2}+h_{2}^{2}).
\ee
With the aid of the equality \eqref{017} it is easy to compute
the optimal signal-to-noise ratio $\rho$ for the signal \eqref{013}:
\be
\label{snr}
\rho := \sqrt{(h|h)}
\cong h_0 \sqrt{\frac{\To}{S_{n}(\fc)}}.
\ee

Substituting Eqs.\ \eqref{017} and \eqref{013} into Eq.\ \eqref{004},
we get the following formula for the likelihood ratio:
\begin{widetext}
\be
\label{018}
\ln\Lambda(x;h_{1},h_{2},\bxi) \cong \frac{2\,\To}{S_{n}(\fc)}\,\bigg(
h_{1}\,\langle x\sin\Phi(t;\bxi)\rangle + h_{2}\,\langle x\,\cos\Phi(t;\bxi)\rangle
- \frac{1}{4} (h_{1}^{2}+h_{2}^{2}) \bigg).
\ee
To maximize $\ln\Lambda$ with respect to the parameters
$h_{1}$ and $h_{2}$ we solve equations
\be
\label{019}
\frac{\partial\ln\Lambda}{\partial h_{i}}=0,\quad i=1,2.
\ee
The unique solution to these equations reads
\be
\label{020}
\hat{h}_{1} = 2\langle x\sin\Phi\rangle,
\quad
\hat{h}_{2} = 2\langle x\cos\Phi\rangle.
\ee
Replacing in Eq.\ \eqref{018} the parameters $h_{1}$ and $h_{2}$
by their estimators $\hat{h}_{1}$ and $\hat{h}_{2}$ [given by Eqs.\ \eqref{020}],
we obtain the reduced likelihood ratio which we call the \emph{$\F$-statistic}:
\be
\label{021}
{\cal F}(x;\bxi) := \ln\Lambda(x;\hat{h}_{1},\hat{h}_{2},\bxi)
\cong \frac{2\,\To}{S_{n}(\fc)} \bigg( \langle x\,\sin\Phi(t;\bxi)\rangle^{2}
+\langle x\,\cos\Phi(t;\bxi)\rangle^{2} \bigg).
\ee
It is rather easy to rewrite the $\F$-statistic in still another form,
\be
\label{021ft}
{\cal F}(x;\bxi) \cong \frac{2}{S_{n}(\fc)\To} \left|\,
\int\limits_{\ti-\To/2}^{\ti+\To/2} x(t) \exp\bigg(-\mi\omega_1\Big(\frac{t}{\To}\Big)^2\bigg)
\exp\Big(-\mi \omega_0 \frac{t}{\To}\Big) \, \md t \, \right|^2.
\ee
\end{widetext}
Thus the $\F$-statistic, up to a constant multiplication factor, is the modulus squared
of the Fourier transform of the product of the data stream $x(t)$
by the exponential factor $\exp\big(-\mi\omega_1\big(t/\To\big)^2\big)$
which depends on the spindown paramerer $\omega_1$.

For the construction of the bank of templates it is crucial to study
the expectation value of the $\F$-statistic \eqref{021} in the case
when the data $x$ contains some gravitational-wave signal $h$, i.e.
\be
x(t) = n(t) + h(t;\btheta'),
\ee
where $\btheta'=(h_1',h_2',\bxi')$ collects the parameters
of the gravitational-wave signal present in the data [see Eq.\ \eqref{013}].
Let us denote this expectation value by $\text{E}_1$
(the subscript `1' means here that the average is computed in the case
when the data contains some gravitational-wave signal),
so we have
\be
\label{E1F}
\text{E}_1\{\F(x;\bxi)\}
= \text{E}\{\F(n(t) + h(t;\btheta');\bxi)\}.
\ee
We want to obtain an approximate analytical formula for this quantity.
Making use of the following approximations
\be
\langle\sin[\Phi(t;\bxi)+\Phi(t;\bxi')]\rangle \cong 0,
\:
\langle\cos[\Phi(t;\bxi)+\Phi(t;\bxi')]\rangle \cong 0,
\ee
after some computation we obtain
\begin{align}
\label{E1Fb}
\text{E}_1\{\F(x;\bxi)\}
\cong 1 + \frac{1}{2} \, \rho^2 \,
\Big( &\big\langle\sin\big[\Phi(t;\bxi)-\Phi(t;\bxi')\big]\big\rangle^{2}
\nonumber\\[1ex]\quad
+ &\big\langle\cos\big[\Phi(t;\bxi)-\Phi(t;\bxi')\big]\big\rangle^{2} \Big),
\end{align}
where $\rho$ is the signal-to-noise ratio from Eq.\ \eqref{snr}.
The right-hand side of the above equation can be rewritten
in terms of the autocovariance function $C_0$ of the $\F$-statistic
(computed in the case when the data contains only noise);
it is defined as
\be
\label{022}
C_0(\bxi,\bxi') := \mathrm{E}\big\{[{\cal F}(n;\bxi)-m_0(\bxi)]
[{\cal F}(n;\bxi')-m_0(\bxi')]\big\},
\ee
where $m_0$ is the signal-free average of $\F$:
\be
\label{023}
m_0(\bxi) := \mathrm{E}\{{\cal F}(n;\bxi)\}.
\ee
In Sec.\ IV of Ref.\ \cite{JK00} it was shown
that the autocovariance function $C_0$
computed for the gravitational-wave signal of the form \eqref{013}
can be approximated by
\begin{align}
\label{C0a}
C_0(\bxi,\bxi') &\cong \big\langle\sin\big[\Phi(t;\bxi)-\Phi(t;\bxi')\big]\big\rangle^{2}
\nonumber\\[1ex]&\quad
+ \big\langle\cos\big[\Phi(t;\bxi)-\Phi(t;\bxi')\big]\big\rangle^{2},
\end{align}
therefore the expectation value \eqref{E1Fb} can shortly be written as
\be
\label{E1Fc}
\text{E}_1\{\F(x;\bxi)\}
\cong 1 + \frac{1}{2} \, \rho^2 \, C_0(\bxi,\bxi').
\ee
The phase $\Phi$ [see Eq.\ \eqref{008}] of the gravitational-wave signal \eqref{013}
depends linearly on the parameters $\bxi$,
therefore the autocovariance function \eqref{C0a} depends only
on the differences between the parameters $\bxi$ and $\bxi'$:
\be
\label{027}
C_0(\bxi,\bxi') \cong \langle\sin\Phi(t;\bxi-\bxi')\rangle^{2}
+ \langle\cos\Phi(t;\bxi-\bxi')\rangle^{2}.
\ee
If one introduces $\btau:=\bxi-\bxi'$, one can thus write
\be
\label{028}
C_0(\btau) \cong \langle\cos\Phi(t;\btau)\rangle^{2}
+ \langle\sin\Phi(t;\btau)\rangle^{2}.
\ee
Let us note that the function $C_0$ attains its maximal value equal to 1
for $\btau=\mathbf{0}$.

We will numerically compute the autocovariance function \eqref{028} in two ways.
First, the right-hand side of Eq.\ \eqref{028} can be expressed
(without any additional approximations)
in terms of the Fresnel integrals:
\begin{widetext}
\begin{align}
\label{035}
\Ce(\btau,\chii) &= \frac{\pi}{2\,|\omega_{1}|}
\Bigg( \bigglb[
\Fc\bigg(\frac{\omega_{0}+\omega_{1}\,(2\chii+1)}{\sqrt{2\,\pi\,|\omega_{1}|}}\bigg)
- \Fc\bigg(\frac{\omega_{0}+\omega_{1}\,(2\chii-1)}{\sqrt{2\,\pi\,|\omega_{1}|}}\bigg)
\biggrb]^2
\nonumber\\[2ex]&\quad + \bigglb[
\Fs\bigg(\frac{\omega_{0}+\omega_{1}\,(2\chii+1)}{\sqrt{2\,\pi\,|\omega_{1}|}}\bigg)
- \Fs\bigg(\frac{\omega_{0}+\omega_{1}\,(2\chii-1)}{\sqrt{2\,\pi\,|\omega_{1}|}}\bigg)
\biggrb]^2 \Bigg),
\end{align}
\end{widetext}
where we have introduced the dimensionless variable
\be
\label{032}
\chii := \frac{\ti}{\To}.
\ee
The Fresnel integrals are defined as\footnote{Let us note
that both $\Fc$ and $\Fs$ are \textit{odd} functions:
$\Fc(-x)=-\Fc(x)$ and $\Fs(-x)=-\Fs(x)$.}
\bse
\label{035a}
\begin{align}
\Fs(x) &:= \int_{0}^{x} \sin\Big(\frac{\pi\,z^2}{2}\Big)\,\md z,
\\[2ex]
\Fc(x) &:= \int_{0}^{x}\cos\Big(\frac{\pi\,z^2}{2}\Big)\,\md z.
\end{align}
\ese

We can also compute the right-hand side of Eq.\ \eqref{028} in an approximate way.
To do this we expand \eqref{028} in Taylor series around $\btau=\mathbf{0}$
up to terms quadratic in $\btau$. Making use of the obvious equalities
\be
\Phi(t;\bxi=\mathbf{0}) = 0,
\quad
\frac{\partial^{2}\Phi}{\partial\xi_{i}\partial\xi_{j}} = 0,
\ee
we get
\be
\label{029}
\Ca(\btau,\chii)\cong 1-\sum_{k,l=1}^2\tilde{\Gamma}(\chii)_{kl}\,\tau_{k}\,\tau_{l},
\ee
where $\tilde{\Gamma}$ is the 2-dimensional
\emph{reduced Fisher information matrix} with elements equal to
\be
\label{030}
\tilde{\Gamma}_{kl} := \Big\langle\frac{\partial\Phi}{\partial\tau_{k}}
\frac{\partial\Phi}{\partial\tau_{l}}\Big\rangle -
\Big\langle\frac{\partial\Phi}{\partial\tau_{k}}
\Big\rangle \Big\langle\frac{\partial\Phi}{\partial\tau_{l}}
\Big\rangle, \quad k,l=1,2.
\ee
In terms of the dimensionless variable $\chii$
the Fisher matrix $\tilde{\Gamma}$ equals
\be
\label{034}
\tilde{\Gamma}(\chii) = \begin{pmatrix}
\frac{1}{12} &
\frac{1}{6}\,\chii
\\[1ex]
\frac{1}{6}\,\chii &
\frac{1}{180}+\frac{1}{3}\,\chii^{2}
\end{pmatrix}.
\ee

\begin{figure*}
\begin{center}
\begin{tabular}{lr}
\includegraphics[scale=0.4]{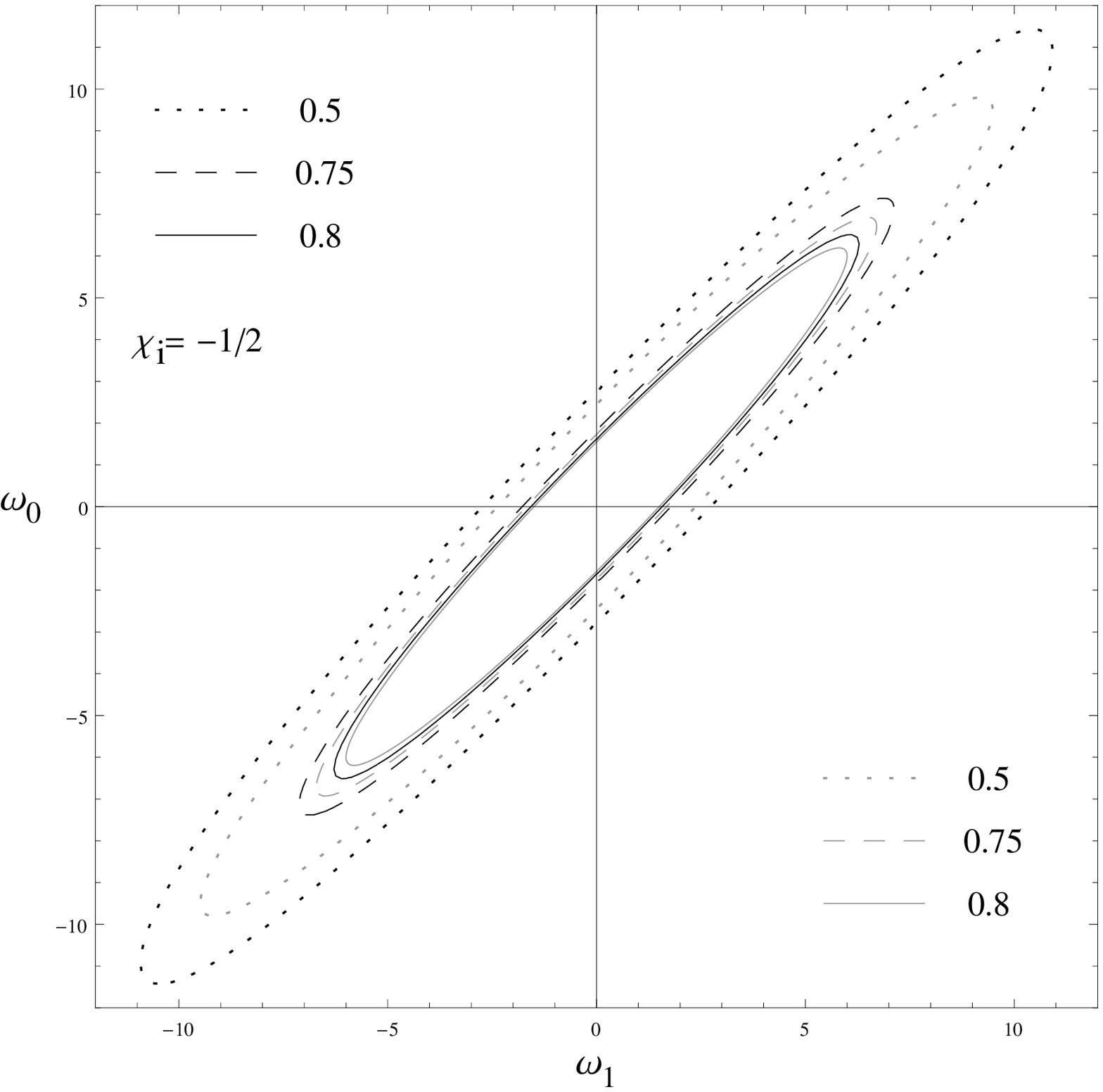} & \includegraphics[scale=0.4]{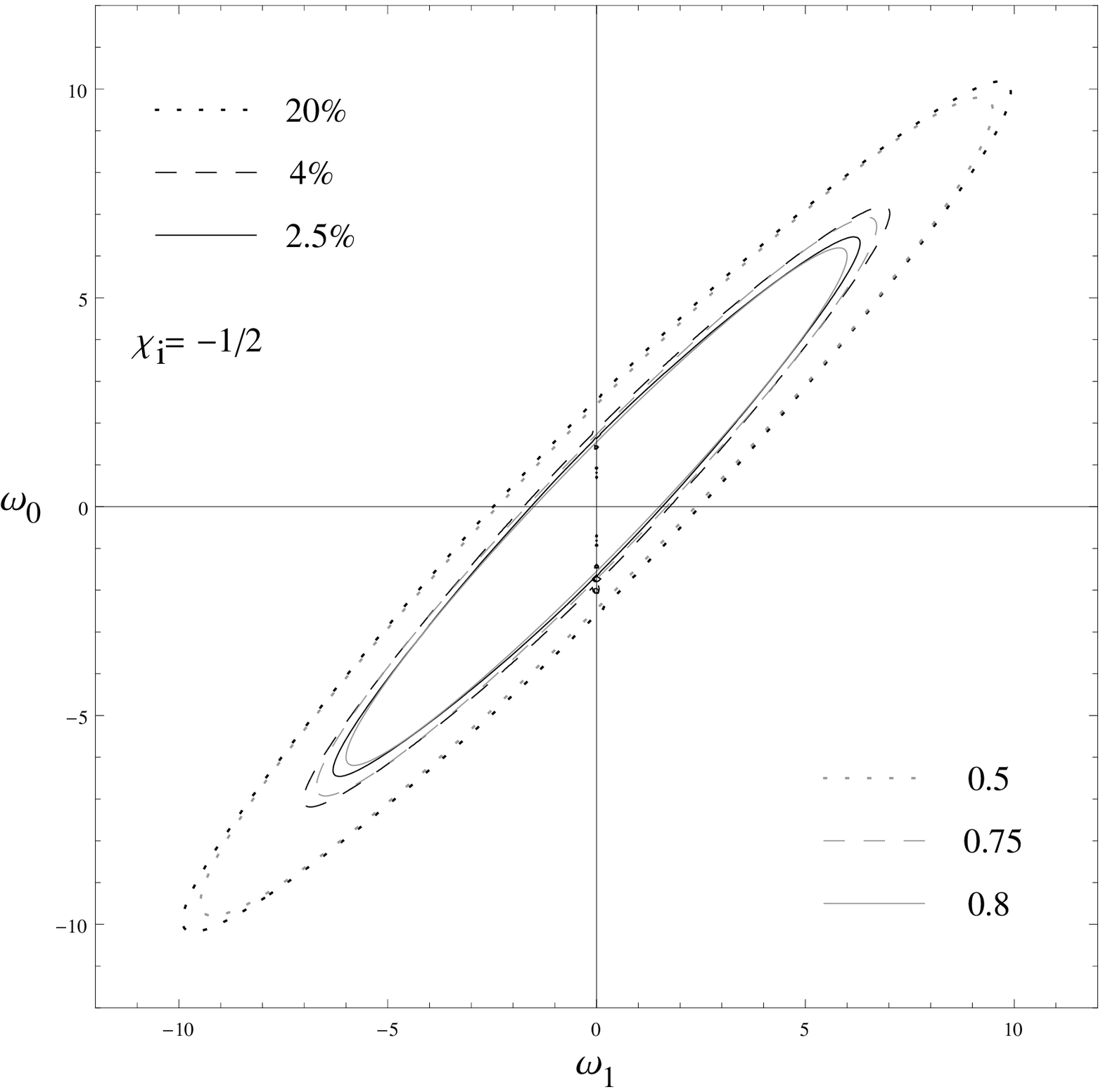}\\
\includegraphics[scale=0.4]{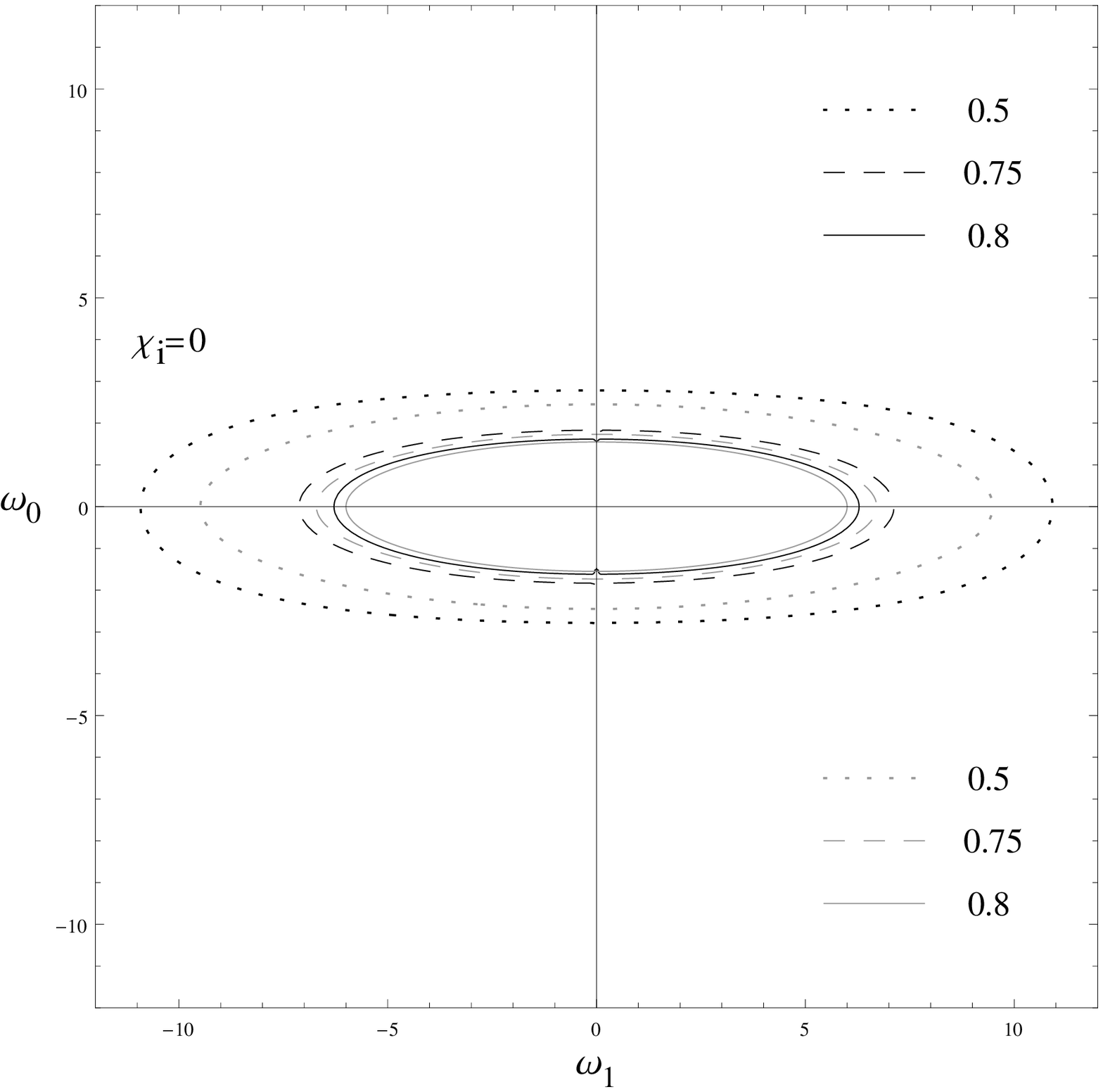} & \includegraphics[scale=0.4]{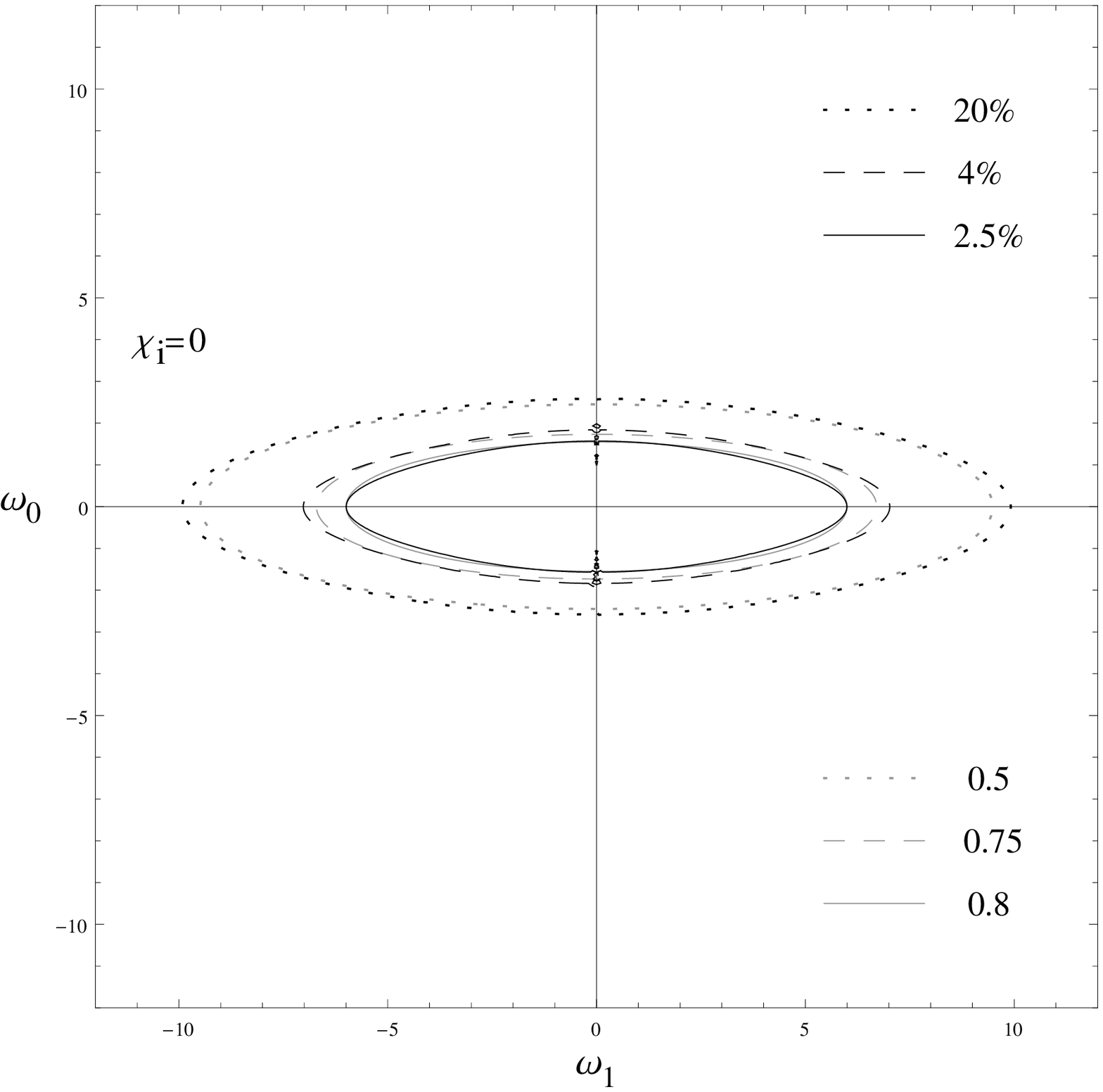}\\
\includegraphics[scale=0.4]{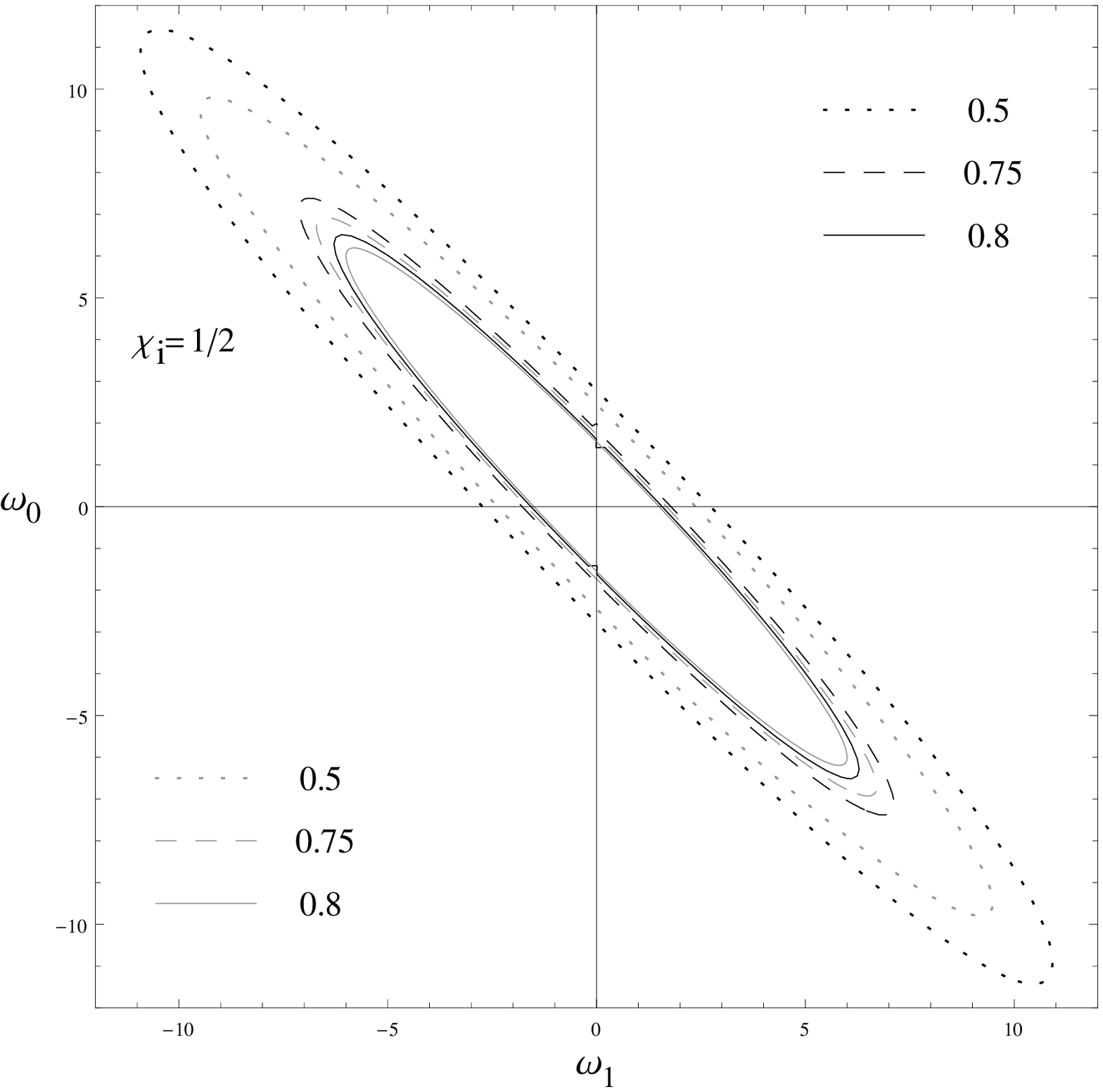} & \includegraphics[scale=0.4]{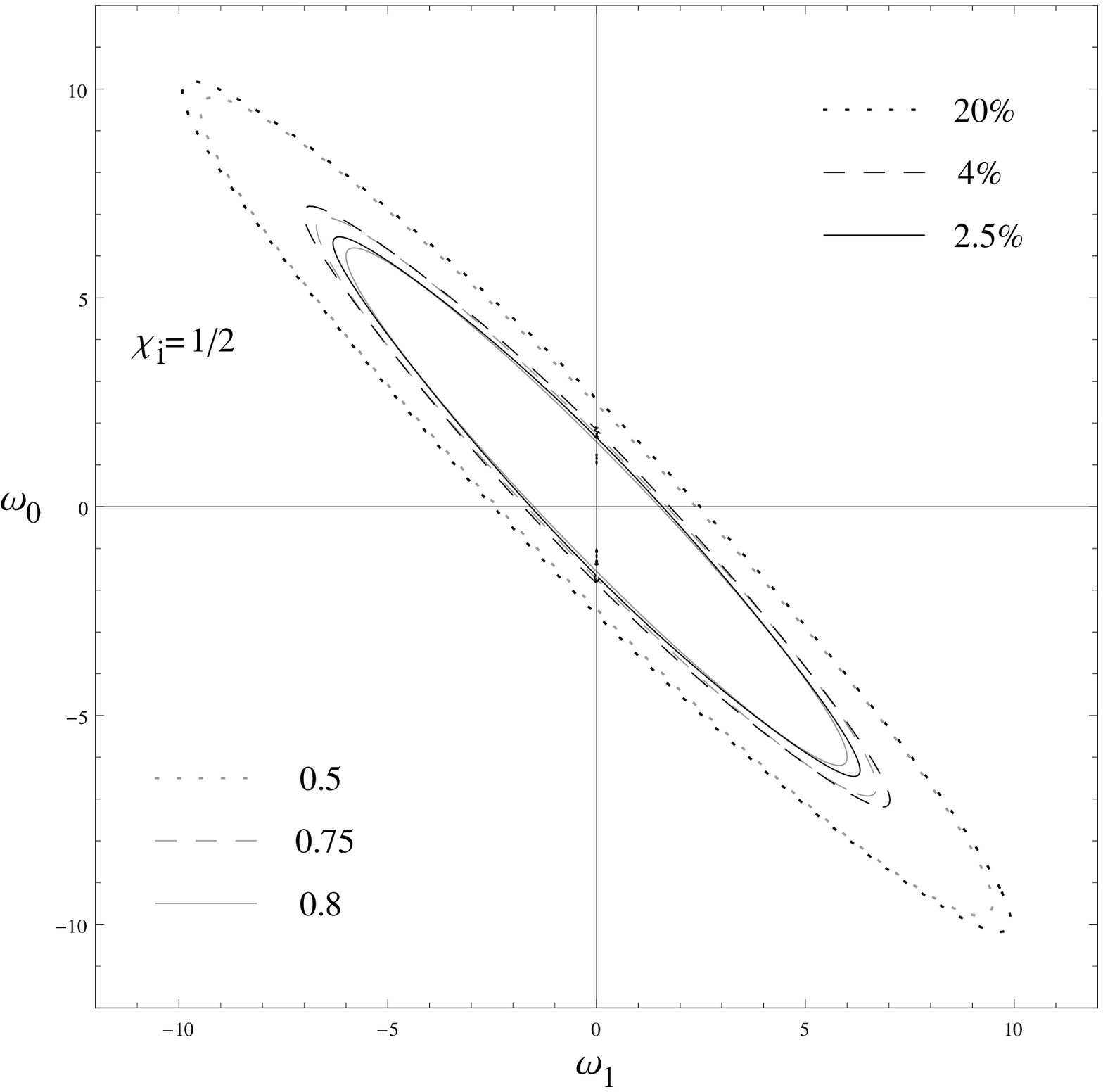}
\end{tabular}
\caption{\label{isoheights}
Left panel: isoheights of the autocovariance function computed by means of the exact formula
\eqref{035} (black lines) and the approximate formula \eqref{029} (grey lines);
isoheights for the values 0.5, 0.75, and 0.8 are shown.
Right panel: isoheights of the autocovariance function computed by means of the approximate formula
\eqref{029} (grey lines) and isoheights of the fractional difference $\Delta C$ defined in Eq. \eqref{036}
(black lines).}
\end{center}
\end{figure*}

In Fig.\ \ref{isoheights} we study the relation between
the exact \eqref{035} and approximate \eqref{029} formulae for the
autocovariance function. We have found (see the left panel of Fig.\ \ref{isoheights})
that the approximate formula underestimates the value of the autocovariance function.
In the right panel of Fig.\ \ref{isoheights}
we plot the isoheights of the approximate autocovariance function
and the isoheights of the fractional difference
\be
\label{036}
\Delta C := \frac{\Ce-\Ca}{\Ce} 100\%.
\ee
One can see that always $\Delta C>0$
and for $\Ce\ge0.75$ the fractional difference $\Delta C<4\%$.

{From} Fig.\ \ref{isoheights} one can also see that whereas
the isoheights of the approximate autocovariance $\Ca$ are perfect ellipses,
the isoheights of the exact autocovariance $\Ce$
are closed curves of shapes very similar to that of ellipses.
Therefore it is reasonable to study the value of the approximate autocovariance
along the isoheight $\Ce=\mathrm{const}$ of the exact autocovariance.
We have done this for the several values of $\Ce$.
Along each $\Ce=\mathrm{const}$ curve the values of the approximate autocovariance
are smaller than $\Ce$ and they are almost the same.
We have picked up the largest value out of them
and these values are given in Table \ref{table1}
together with the corresponding values of $\Ce$
and the fractional difference between $\Ca$ and $\Ce$.
The following cubic fit,
\be
\Ce = -0.823872 + 3.15065\,\Ca - 1.84055\,\Ca^2 + 0.514037\,\Ca^3,
\ee
reproduces the relation $\Ce=\Ce(\Ca)$ with accuracy better than 0.04\%
for $0.55\le\Ca\le1.00$.

\begin{table}
\caption{\label{table1}
The relation between the exact, Eq.\ \eqref{035}, and the approximate, Eq.\ \eqref{029},
formulae of the autocovariance function of the $\F$-statistic.}
\begin{ruledtabular}
\begin{tabular}{llr}
$\Ce$ & $\Ca$ & $(\Ce-\Ca)/\Ce$\\ \hline
  0.55 & 0.43767 & 20.4\% \\
  0.60 & 0.51513 & 14.2\% \\
  0.65 & 0.58769 &  9.6\% \\
  0.70 & 0.65600 &  6.3\% \\
  0.75 & 0.72057 &  3.9\% \\
  0.80 & 0.78183 &  2.3\% \\
  0.85 & 0.84012 &  1.2\% \\
  0.90 & 0.89575 &  0.5\% \\
  0.95 & 0.94897 &  0.1\% \\
\end{tabular}
\end{ruledtabular}
\end{table}

\section{Banks of the templates}

To search for the gravitational-wave signal \eqref{010} in detector's noise
we need to construct a bank of templates in the space of the parameters $(\omega_0,\omega_1)$
on which the $\F$-statistic [given in Eq.\ \eqref{021}] depends.
The bank of templates is defined by a discrete set of points, i.e.\ a \emph{grid} in the parameter space
chosen in such a way, that for any possible signal there exists a grid point
such that the expectation value of the $\F$-statistic computed for the parameters
of this grid point is not less than a certain fixed minimal value.
From Eq.\ \eqref{E1Fc} we see that this expectation value
depends on the signal-to-noise ratio $\rho$
and on the value of the noise autocovariance function $C_0$
computed for the intrinsic parameters $\bxi$ and $\bxi'$
of the template and the gravitational-wave signal, respectively.
\emph{In the rest of this paper we will approximate the autocovariance function $C_0$
by means of the formula \eqref{029}},
i.e.\ we will use the equality $C_0(\bxi,\bxi')\cong\Ca(\bxi,\bxi')$.

The signal-to-noise ratio $\rho$ we can not control,
therefore to construct the bank of templates
one needs to choose some minimum value $\cmin$ of the autocovariance function $\Ca$
and look for such a grid of points that for any point $\bxi'$ in the $(\omega_0,\omega_1)$ plane
there exists a grid node $\bxi$ such that the autocovariance $\Ca$ computed for the parameters
$\bxi$ and $\bxi'$ is not less than $\cmin$.
Because $\Ca$ depends on $\bxi$, $\bxi'$ only through the difference $\bxi-\bxi'$,
we require that
\be
\label{ellipse1}
\Ca(\bxi-\bxi') \ge \cmin.
\ee
By virtue of Eq.\ \eqref{029} this condition leads to the inequality
\be
\label{ellipse2}
\sum_{k,l=1}^2\tilde{\Gamma}_{kl}\,(\xi_k-\xi'_k)\,(\xi_l-\xi'_l)
\le 1 - \cmin,
\ee
which for the fixed $\bxi$ is fulfilled by all points $\bxi'$ which belong
to an ellipse with the center located at $\bxi$.

Wa want to find the \emph{optimal} grid fulfilling the requirement \eqref{ellipse1},
i.e.\ the grid which consists of possibly smallest number of points.
Thus the problem of finding the optimal grid is a kind of \emph{covering} problem,
i.e.\ the problem to cover the $(\omega_0,\omega_1)$ plane (or, in data analysis
case, the bounded region of the plane) by the smallest number of \emph{identical} ellipses.
The thorough exposition of the problem of covering $n$-dimensional Euclidean space
by identical spheres is given in Chap.\ 2 of Ref. \cite{CS99}.

We restrict ourselves to grids which are \emph{lattices},
i.e.\ to grids with nodes which are linear combinations
(with integer coefficients) of two basis vectors.
If the vectors ($\mathbf{P}_0$, $\mathbf{P}_1$) are the basis vectors of a lattice,
then a \emph{fundamental parallelogram} (for the case of more than two parameters
it would be a parallelotope) is the parallelogram consisting of the points
\be
\lambda_0 \mathbf{P}_0 + \lambda_1 \mathbf{P}_1, \quad
0 \le \lambda_0,\lambda_1 \le 1.
\ee
A fundamental parallelogram is an example of a \emph{fundamental region}
for the lattice, which when repeated many times fills the plane
with one lattice point in each copy.
The quality of a covering can be expressed by the \emph{covering thickness} $\theta$
which is defined as the average number of ellipses that contain a point in the plane.
For lattice coverings their thickness can be computed as
\be
\theta = \frac{\text{area of one ellipse}}{\text{area of fundamental region}}.
\ee

We assume that we are interested in such searches for almost monochromatic gravitational-wave signals
for which the number of grid points in the parameter space is very large
and the time needed to compute the $\F$-statistic for all grid nodes is long,
so it is crucial to use in the computation the fast numerical algorithms.
Because the computation of the $\F$-statistic involves calculation of the Fourier transform
[see Eq.\ \eqref{021ft}], one would like to use the \emph{fast Fourier transform} (FFT) algorithm.
The FFT algorithm computes the values of the \emph{discrete Fourier transform} (DFT)
of a time series. The values of the DFT are defined for a certain set of discrete frequencies
called the \emph{Fourier frequencies}.
It will thus be possible to use the FFT algorithm
in computation of the $\F$-statistic,
if the grid points will be arranged in such a way,
that the frequency coordinates of these points
will all coincide with the Fourier frequencies.
We have constructed two different families of grids which fulfill this requirement,
details of the construction are presented in Appendix \ref{grids}.
In the construction we have explored observation
that the shape of the autocovariance ellipse
strongly depends on the value of the parameter $\chii$:
the larger $|\chii|$ is, the more alongated (along the $\omega_0$-axis) this ellipse is
(see Fig.\ \ref{isoheights}).
But the usage of alongated enough ellipses suggests that the $\F$-statistic could be computed
for smaller number of frequency values than the standard DFT algorithm computes.
Therefore we have employed some modifications of the DFT algorithm.

Le the data collected by a detector form a sequence of $N$ samples
\be
x_u, \quad u=1,\ldots,N,
\ee
and let the sampling period be $\Delta{t}$. Then the DFT algorithm calculates
the Fourier transform of the data with the frequency resolution $\Delta{f}=1/(N\Delta{t})$.
The resolution of the dimensionless frequency parameter $\omega_0$
[introduced in Eq.\ \eqref{007c}] is thus
\be
\label{Domega0}
\Delta{\omega_0} = 2\pi\To \Delta{f} = 2\pi,
\ee
because $N\Delta{t}=\To$.
It is possible to modify the DFT algorithm in such a way,
that the frequency resolution \eqref{Domega0} changes.
In Appendix \ref{aDFT} we consider two such modifications:
(i) zero-padding of the data, which makes the DFT more dense;
(ii) folding of the data, which diminishes the frequency resolution.
Therefore we study banks of templates
which are compatible with the frequency resolutions of the form
\be
\label{Domega0m}
\Delta{\omega_0} = 2^\ell\pi, \quad \ell=0,1,2,3.
\ee
As explained in Appendix \ref{aDFT},
$\ell=0$ corresponds to zero-padding of the data
(for $N$ data points we add $N$ zeros),
$\ell=1$ is a pure DFT of $N$-point data stream,
for $\ell=2$ the data is folded two times
and for $\ell=3$ the data is folded four times.

Let ($\mathbf{P}_0$, $\mathbf{P}_1$) be the basis vectors of a lattice we consider.
As explained above we want to use the DFT algorithm,
therefore we need a such bank of templates
that all nodes can be arranged along straight lines parallel to the $\omega_{0}$-axis.
Moreover, the distance between neighboring nodes along these lines
must be equal to the frequency resolution \eqref{Domega0m} of the DFT algorithm.
To fulfill this constraint we require that the vector $\mathbf{P}_0$ has components
\be
\label{constraint}
\mathbf{P}_0 = (\Delta{\omega_0}=2^\ell\pi,\,0).
\ee
Let us also denote the components of the second basis vector $\mathbf{P}_1$ as
\be
\label{baseP1}
\mathbf{P}_1 =(\delta\omega_0,\delta\omega_1).
\ee
We will call the grid \emph{orthogonal}, if $\mathbf{P}_0\cdot\mathbf{P}_1=0$,
where dot denotes the usual Euclidean scalar product.
The grid spanned by the vectors \eqref{constraint} and \eqref{baseP1} is thus orthogonal
if and only if $\delta\omega_0=0$.

In Appendix \ref{grids} we have constructed two different families of grids
fulfilling the constraint \eqref{constraint}.
Construction of the grids denoted by G$_{1,\ell}$ (for $\ell=0,1,2,3$)
is described in Appendix \ref{grid1},
the grids G$_{2,\ell}$ and G$'_{2,\ell}$ (valid for $\ell=1,2,3$)
are constructed in Appendix \ref{grid2},
and the grids G$_{2,0}$ and G$'_{2,0}$ are described in Appendix \ref{grid3}.
The grids G$_{1,\ell}$ and G$'_{2,\ell}$ are orthogonal,
whereas the grids G$_{2,\ell}$ are non-orthogonal.

\section{Time performance of the grids}

We have made a number of numerical simulations to study the performance
of the grids constructed in Appendix \ref{grids}.
All computations were done for $N=2^{19}=524288$ data points.
For the sampling period $\Delta{t}=0.5$~s (i.e.\ for the Nyquist frequency equal to 1~Hz)
this corresponds to around three days of data.
The dimensionless first spindown parameter $\omega_1$
we have taken to be nonpositive and in the range $\omega_1\in\langle-3000;0\rangle$.
The discrete-in-time version of the $\F$-statistic
[its continuous-in-time form is given in Eq.\ \eqref{021ft}] reads
\begin{widetext}
\be
\label{Fdis1}
{\cal F}(x;\bxi) \cong \frac{2\Delta{t}^2}{S_{n}(\fc)\To} \left|\,
\sum_{u=1}^N x_u \exp\left[-\mi\omega_1 \left(\chii-\frac{1}{2}+\frac{u-1}{N}\right)^2\right]
\exp\left(-\mi\omega_0\frac{u-1}{N}\right) \right|^2,
\ee
\end{widetext}
where the discrete data stream points are defined as $x_u:=x\big(\ti-\To/2+(u-1)\Delta{t}\big)$
($u=1,\ldots,N$). The DFT algorithm computes the sum present in Eq.\ \eqref{Fdis1}
simultaneously for all dimensionless frequencies $\omega_0$ from a discrete set.
Let us denote these discrete Fourier frequencies by $\omega_{0r}$, then
(see Appendix A)
\be
\label{omega0r}
\omega_{0r} = (r-1)\,\Delta\omega_0, \quad r=1,\ldots,r_\text{max},
\ee
where $\Delta\omega_0$ is the frequency resolution of the DFT.
Using in the parameter space a discrete grid
spanned by the vectors $(\mathbf{P}_0,\mathbf{P}_1)$
[see Eqs.\ \eqref{constraint} and \eqref{baseP1}]
means that also the spindown parameter $\omega_1$
becomes discrete with possible values equal to
\be
\label{omega1s}
\omega_{1s} = -(s-1)\,\delta\omega_1, \quad s=1,\ldots,s_\text{max}.
\ee

In the case of non-orthogonal grids the point in the parameter space
with coordinates $(\mbox{$\omega_0=0$},\omega_{1s})$
usually is not a grid node for $s\ne0$;
the grid node with the smallest $\omega_0$-coordinate is
$(\text{mod}(s\,\delta\omega_0,\Delta\omega_0),\omega_{1s})$,
where $\text{mod}(m,n)$ is the remainder on division of $m$ by $n$.
It means that for the fixed value of $s\ne0$,
before computing the DFT one has to multiply the $u$th data point by an extra exponential factor
$\exp[-\mi\,\text{mod}(s\,\delta\omega_0,\Delta\omega_0)u/N]$.

Collecting all this information together
one can finally show that the $\F$-statistic evaluated on the grid point defined
by the integers $(r,s)$ [introduced in Eqs.\ \eqref{omega0r} and \eqref{omega1s}]
can be written as
\begin{widetext}
\begin{align}
\label{Fdis2}
\F(x;r,s) \cong \frac{2\Delta{t}^2}{S_{n}(\fc)\To} \bigg|\,
\sum_{u=1}^N x_u & \exp\left[ \mi\,(s-1)\,\delta\omega_1 \left(\chii-\frac{1}{2}+\frac{u-1}{N}\right)^2
-\mi\,\text{mod}(s\,\delta\omega_0,\Delta\omega_0)\frac{u}{N} \right]
\nonumber\\[1ex]
\times & \exp\left[-\mi\,(r-1)\,\Delta\omega_0\frac{u-1}{N}\right] \bigg|^2,
\quad r=1,\ldots,r_\text{max},\:s=1,\ldots,s_\text{max}.
\end{align}
\end{widetext}

\begin{table*}
\caption{\label{simula}
The results of numerical simulations showing the time performance
of the grids constructed in Appendix \ref{grids}.
Besides the times needed to compute the $\F$-statistic
[given in Eq.\ \eqref{Fdis2}] for all grid nodes,
the number of the FFTs calculated during the computation
and the covering thicknesses $\theta$ are also shown.
All grids are constructed for $\cmin=0.75$ (which corresponds to $\cmine=0.7737$).
Each time slot in the table is the arithmetic mean from 200 repetitions
of the computation of the $\F$-statistic for all grid nodes.}
\begin{ruledtabular}
\begin{tabular}{ccccccccc}
& \multicolumn{3}{c}{G$_{1,\ell}$}
& \multicolumn{5}{c}{G$_{2,\ell}$, G$'_{2,\ell}$, G$''_{2,\ell}$} \\
\cline{2-4}\cline{5-9}\multirow{2}{*}{$\ell$}
& \multirow{2}{*}{Time (s)}
& \multirow{2}{*}{No of FFTs}
& \multirow{2}{*}{$\theta$}
& \multicolumn{3}{c}{Time (s)}
& \multirow{2}{*}{No of FFTs}
& \multirow{2}{*}{$\theta$} \\ \cline{5-7}
& & & & G$_{2,\ell}$ & G$'_{2,\ell}$ & G$''_{2,\ell}$ &  & \\ \hline
0 & 23.4 &  318 & 1.2287 &  48.1 & 23.1 & 30.4 &  315 & 1.2186 \\
1 & 31.3 &  705 & 1.3639 &  81.7 & 32.2 &   -- &  723 & 1.3990 \\
2 & 38.2 & 1273 & 1.2325 & 126.6 & 38.8 &   -- & 1276 & 1.2351 \\
3 & 63.0 & 2676 & 1.2954 & 230.2 & 59.1 &   -- & 2501 & 1.2106
\end{tabular}
\end{ruledtabular}
\end{table*}

In our numerical simulations we have used the FFT algorithm\footnote{
We have employed the fftw 3.2.2 implementation of the FFT algorithm
with the wisdom mechanism (within which the algorithm performs some tests
to ensure the optimal FFT performance).}
with different frequency resolutions given in Eq.\ \eqref{Domega0m}.
In Table \ref{simula} we have shown the results of simulations
in which we have studied, for different types of grids,
the time needed to compute the $\F$-statistic for all grid nodes
and the number of Fourier transforms performed during the computation.\footnote{
Our codes were written in C and compiled with gcc 4.3.4.
We have used PC computer with Core 2 Quad 2.66 GHz processor and 4 GB of RAM.}
We have also shown the thicknesses $\theta$ of coverings for different grids.

Let us remind that in two dimensions the optimal covering (without
any constraint imposed) is the hexagonal covering with thickness $\theta$
equal to $2\pi/(3\sqrt{3})\cong1.2092$.
The grids presented in Table \ref{simula} have all thicknesses
in the range $1.2106\lesssim\theta\lesssim1.3990$,
so they are greater than the optimal hexagonal covering
thickness by $\sim$0.1\% to $\sim$15.7\%.
{From} inspection of Table \ref{simula} it is clear
that the value of the thickness does not decide
on efficiency of a grid. We see
that the most efficient grids are grids which involve
the computation of the longest FFTs. Moreover, within the grids
with the same value of $\ell$ (i.e.\ with the same length of the FFTs),
much more efficient are orthogonal grids.
The reason for this is the lack of the exponential term
$\exp[-\mi\,\text{mod}(s\,\delta\omega_0,\Delta\omega_0)u/N]$
in the discrete form of the $\F$-statistic, see Eq.\ \eqref{Fdis2}.

The non-orthogonal grid G$''_{2,0}$ is identical to the grid G$_{2,0}$,
but the time-performances of these two grids are different,
because of different way of handling,
during the computation of the $\F$-statistics, the exponential term
$\exp[-\mi\,\text{mod}(s\,\delta\omega_0,\Delta\omega_0)u/N]$ in Eq.\ \eqref{Fdis2}.
The computing time can be considerably reduced if the values
of this term are computed in advance for all needed values of $s$ and $u$
and they are kept in RAM memory of computer (note that the exponential term
depends on the quantities $\Delta\omega_0$ and $\delta\omega_0$ defining the grid
and does not depend on data). However, to apply this trick one has to reserve
a large amount of RAM for the table with values of the exponential term.
For $\ell=0$ (i.e.\ for the grid G$''_{2,0}$) this table requires about 2.5 GB of RAM.
One can estimate that the table for $\ell+1$ requires around 2 times more RAM
than the table for $\ell$.

\appendix

\section{Modifications of the discrete Fourier transform}
\label{aDFT}

The data from the detector form the sequence
\be
\label{data}
(x_r) = (x_1, x_2, \ldots, x_{N}),
\ee
so $N$ is the number of data points.
The discrete Fourier transform (DFT) of the data \eqref{data} is defined as
\be
\label{220}
\tilde{x}_s =\sum\limits_{r=1}^{N} x_r
\exp \bigg(-2\pi\mi\frac{(r-1)(s-1)}{N}\bigg),
\quad s=1,\ldots,N.
\ee
The DFT defined above computes the Fourier transform of the data stream \eqref{data}
at frequencies
\be
\label{220a}
f_s = \frac{s-1}{N\Delta{t}} = 2(s-1)\frac{\fN}{N},
\quad s=1,\ldots,N,
\ee
where $\Delta{t}$ is the sampling period and $\fN$ is the Nyquist frequency.
The frequency resolution of the DFT \eqref{220} is thus
\be
\Delta{f} = \frac{1}{N\Delta{t}} \, \Longrightarrow \, \Delta{\omega_0} = 2\pi.
\ee

\subsection{Zero padding}

Let us now consider the $2N$-point data stream $(y_r)$ which consists
of the original $N$-point data stream $(x_r)$, Eq.\ \eqref{data},
supplemented by $N$ zeros,
\be
\label{zp1}
(y_r) = (x_1,\ldots,x_{N},0,\ldots,0).
\ee
By virtue of formula \eqref{220},
the DFT of the data $(y_r)$ reads
\be
\label{zp2}
\tilde{y}_s =\sum\limits_{r=1}^{N} x_r
\exp \bigg(-2\pi\mi\frac{(r-1)(s-1)}{2N}\bigg),
\quad s=1,\ldots,2N.
\ee
Making use of Eqs.\ \eqref{220} and \eqref{220a} one easily sees,
that the numbers $\tilde{y}_s$ can be interpreted
as the values of the DFT of the original data stream $(x_r)$,
but computed now for frequences
\be
\label{zp3}
f_s = \frac{s-1}{2N\Delta{t}},
\quad s=1,\ldots,2N,
\ee
so the frequency resolution of the DFT \eqref{zp2} is 
\be
\label{zp4}
\Delta{f} = \frac{1}{2N\Delta{t}} \, \Longrightarrow \, \Delta{\omega_0} = \pi.
\ee
Of course one can add more zeros to the data to obtain their DFT
with the frequency resolution better than this given in Eq.\ \eqref{zp4}.

\subsection{Folding of data}

Let us fold the $N$-point data stream $(x_r)$
to the $(N/2)$-point data stream $(y_r)$,
\be
\label{fd1}
y_r := x_r + x_{r+N/2}, \quad r=1,\ldots,\frac{N}{2}.
\ee
The DFT of the data $(y_r)$ reads, according to Eq.\ \eqref{220},
\begin{align}
\label{fd2}
\tilde{y}_s &=\sum\limits_{r=1}^{N/2} (x_r + x_{r+N/2})
\exp \bigg(-2\pi\mi\frac{(r-1)(s-1)}{N/2}\bigg),
\nonumber\\
&\qquad s=1,\ldots,\frac{N}{2}.
\end{align}
It is not difficult, employing the periodicity of the function $\exp(\mi z)$ for real $z$,
to rewrite formula \eqref{fd2} in the form
\begin{align}
\label{fd3}
\tilde{y}_s =\sum\limits_{r=1}^{N} x_r
\exp \bigg(\!-2\pi\mi\frac{2(r-1)(s-1)}{N}\bigg),
\quad s=1,\ldots,\frac{N}{2}.
\end{align}
Again making use of Eqs.\ \eqref{220} and \eqref{220a} one sees,
that the numbers $\tilde{y}_s$ can be interpreted
as the values of the DFT of the original data stream $(x_r)$,
but computed for frequences
\be
\label{fd4}
f_s = \frac{2(s-1)}{N\Delta{t}},
\quad s=1,\ldots,\frac{N}{2},
\ee
so the frequency resolution of the DFT \eqref{fd3} is 
\be
\label{fd5}
\Delta{f} = \frac{2}{N\Delta{t}} \, \Longrightarrow \, \Delta{\omega_0} = 4\pi.
\ee
By folding the data $p$ times ($p=1,2,\ldots$)
one gets $(N/p)$-point data stream
the DFT of which is the DFT of the original $N$-point data
but computed with the frequency resolution $\Delta{\omega_0}=2^p\times2\pi$.

\section{Construction of the grids in the parameter space}
\label{grids}

\subsection{Grids $G_1$}
\label{grid1}

In this appendix we construct a family of grids
by construction of their fundamental regions of parallelogram shape.
The fundamental parallelograms are always inscribed
into the ellipse of constant value of the autocovariance function
[this ellipse is given by Eq.\ \eqref{ellipse2}]
and their construction ensures that the constraint \eqref{Domega0m} is fulfilled.

\subsubsection{Construction of the fundamental parallelogram}
\label{ParaEleCell}

Let us denote the coordinates of the parallelogram vertexes we are looking for
by $(\omega_0^{(a)},\omega_1^{(a)})$, $a=1,\ldots,4$.
Bases of the parallelogram we choose to be parallel to the $\omega_{1}$ axis,
so the $\omega_{0}$ coordinates of the vertexes can be written as
\be
\label{201}
\omega_0^{(a)}(k) = k\,\frac{\pi}{2}, \quad a=1,2, \quad
\omega_0^{(a)}(k) = -k\,\frac{\pi}{2}, \quad a=3,4,
\ee
where $k>0$. Then the coordinates $\omega_{1}^{(a)}$ of the vertexes we obtain from the equations
[here $\mathsf{T}$ denotes the transposition of the row vector $(\omega_0^{(a)},\omega_1^{(a)})$]:
\be
\label{203}
(\omega_{0}^{(a)},\omega_{1}^{(a)})\cdot\tilde{\Gamma}(\chii)\cdot
(\omega_{0}^{(a)},\omega_{1}^{(a)})^\mathsf{T}=1-\cmini, \quad a=1,\ldots,4,
\ee
where $\tilde{\Gamma}(\chii)$ is the Fisher matrix from Eq.\ \eqref{034}
and $\cmini$ (the reason for introducing the extra subscript `i', from `initial',
is explained in Appendix \ref{G1construction} below)
is the minimum value of the autocovariance function.
We can assume that $\cmini\in(0;\,1)$, then the solution of Eqs.\ \eqref{203} reads
\begin{widetext}
\bse
\label{204}
\begin{align}
\label{204a}
\omega_{1}^{(1)}(\chii,k) &= \frac{-30\,k\,\pi\,\chii +
\sqrt{15}\sqrt{48\,(1-\cmini)\,(1+60\,\chii^{2})-k^2\,\pi^2}}{2\,(1+60\,\chii^{2})},
\\[2ex]
\label{204b}
\omega_{1}^{(2)}(\chii,k) &= \frac{-30\,k\,\pi\,\chii -
\sqrt{15}\sqrt{48\,(1-\cmini)\,(1+60\,\chii^{2})-k^2\,\pi^2}}{2\,(1+60\,\chii^{2})},
\\[2ex]
\label{204cd}
\omega_{1}^{(3)}(\chii,k) &= -\omega_{1}^{(2)}(\chii,k),
\qquad
\omega_{1}^{(4)}(\chii,k) = -\omega_{1}^{(1)}(\chii,k).
\end{align}
\ese
\end{widetext}
Area of the parallelogram with the vertexes given above is equal to
\be
\label{205a}
S=(\omega_{1}^{(1)}-\omega_{1}^{(2)})\,k\,\pi.
\ee
Making use of Eqs.\ \eqref{204} we get the area $S$
as a function of the parameters $\chii$ and $k$:
\be
\label{205b}
S(\chii,k) = \frac{\sqrt{15}\sqrt{48\,(1-\cmini)\,(1+60\,\chii^{2})-k^2\,\pi^2}}
{1+60\,\chii^{2}}\,k\,\pi.
\ee

We want to have the fundamental parallelogram of possibly large area.
Let us first fix the parameter $\chii$ and maximize the area with respect to the parameter $k$.
The only positive solution of the equation $\partial S/\partial k=0$ reads
\be
\label{207}
k_0(\chii) = \frac{2}{\pi}\,\sqrt{6\,(1-\cmini)(1+60\,\chii^{2})}.
\ee
We can compute the area of the parallelogram for this optimal value of $k$:
\be
\label{208}
S_{\mathrm{max}} = S\big(\chii,k_0(\chii)\big)= 24\sqrt{15}\,(1-\cmini).
\ee
It does not depend on the vaule of $\chii$.
It can be shown that the value $S_\mathrm{max}$ given above
is also the \emph{global} maximum of the function $S=S(\chii,k)$.

\begin{figure*}
\begin{center}
\begin{tabular}{lr}
\includegraphics[scale=0.55]{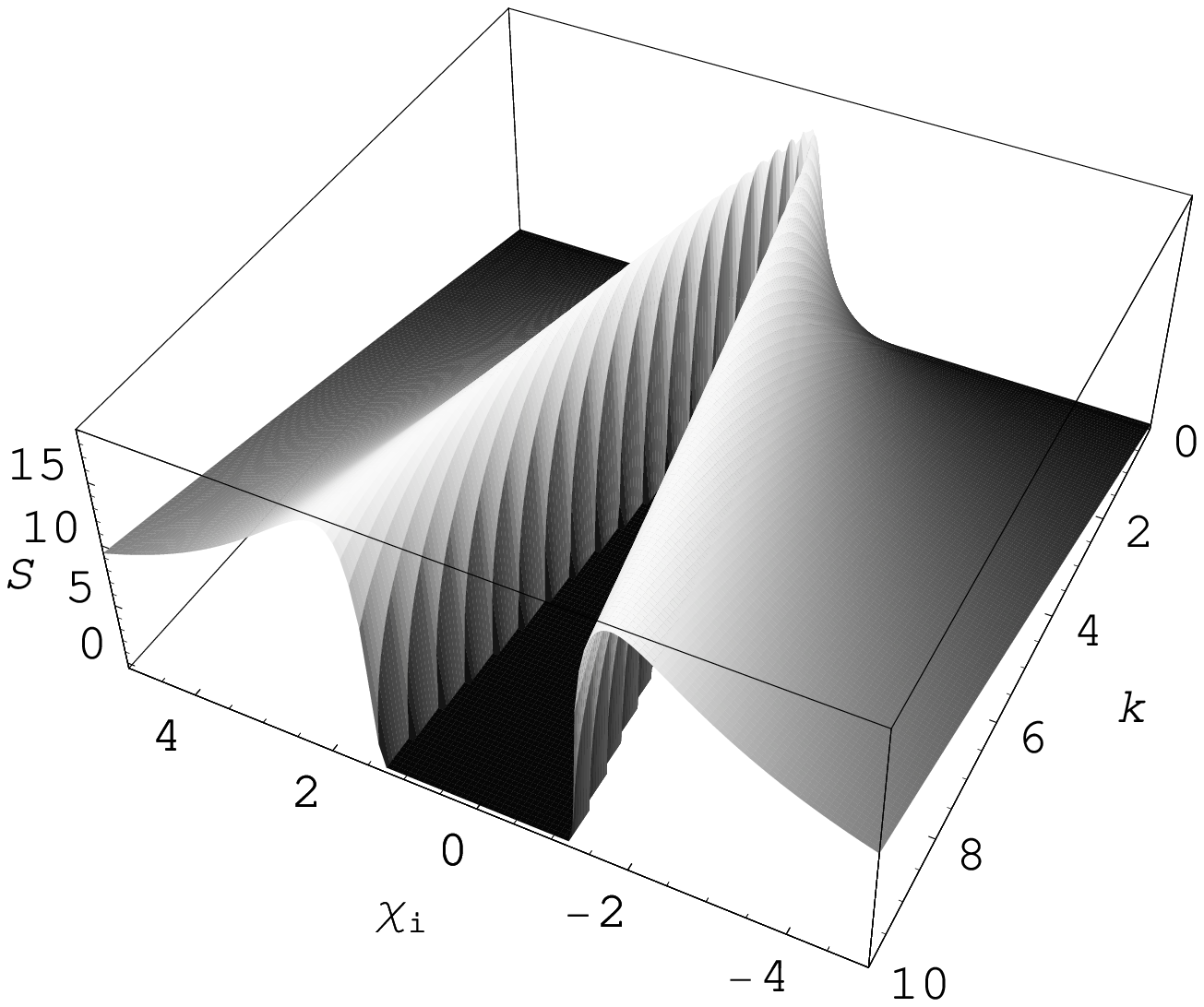}\qquad
& \qquad\includegraphics[scale=0.50]{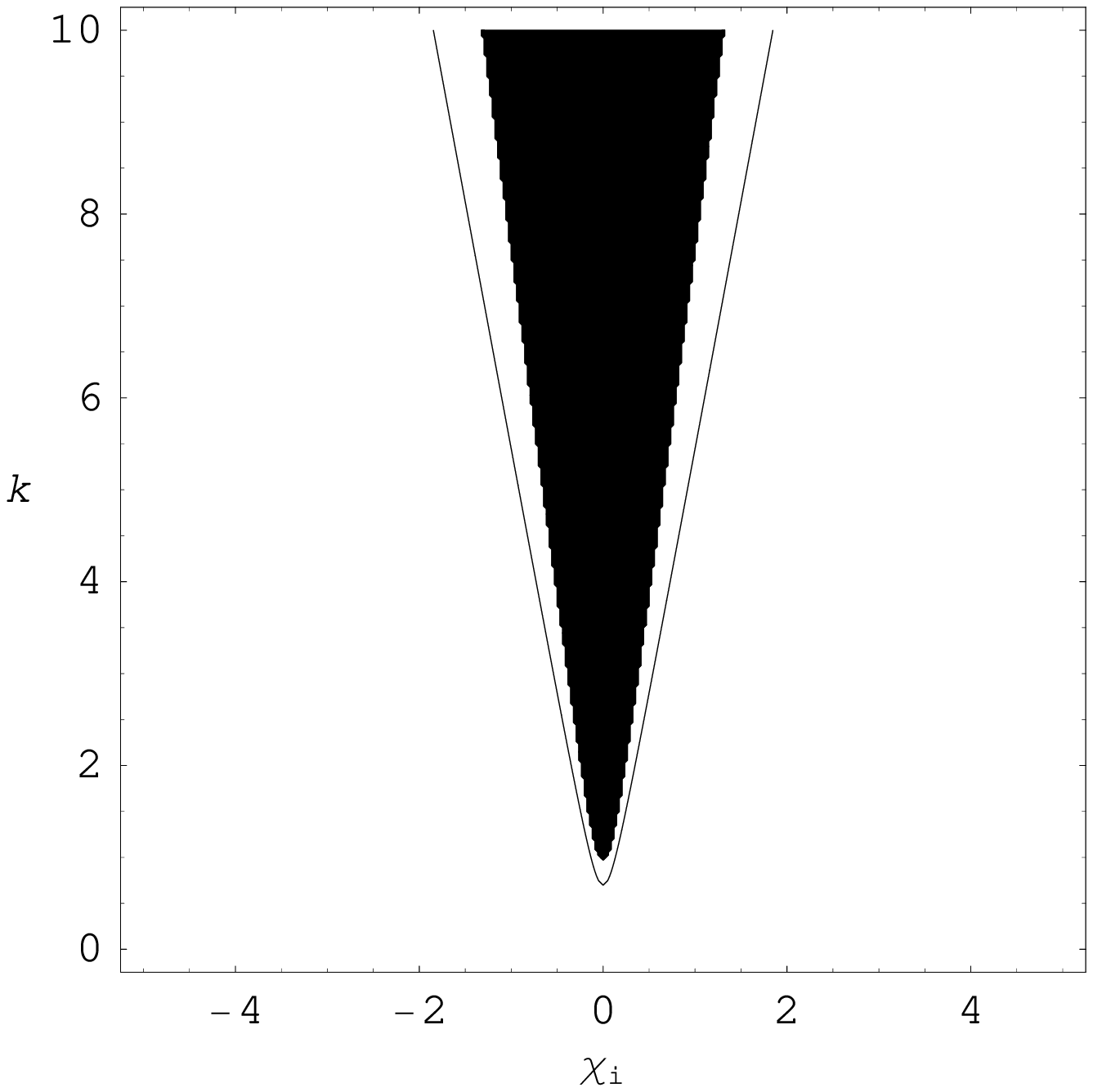}
\end{tabular}
\caption{\label{area}
In the left panel the area $S$ of the fundamental parallelogram for the grid $G_1$
is plotted as a function of the parameters $\chii$ and $k$ [see Eq.\ \eqref{205b}].
In the right panel the part of the $(\chii,k)$ plane is displayed; here the shaded region
is made of the points $(\chii,k)$ for which the construction of the parallelogram is not possible,
and the V-shaped solid line is made of these values of the parameters $(\chii,k)$
for which the maximal area $S_{\mathrm{max}}$ is achieved.}
\end{center}
\end{figure*}

The dependence $S=S(\chii,k)$ is illustrated in Fig.\ \ref{area}.
It is easy to show that for any value of $k>k_{\mathrm{min}}$, where
\be
\label{211}
k_{\mathrm{min}}:=\frac{2}{\pi}\sqrt{6\,(1-\cmini)},
\ee
we can find two values of $\chii$ for which the maximal area $S_{\mathrm{max}}$ is achieved.
These values one obtains solving the equation $S(\chii,k)=S_{\mathrm{max}}$. The solutions read
\be
\label{209}
\chi_{\mathrm{i}}(k) =
\pm \frac{\sqrt{k^2\,\pi^2-24(1-\cmini)}}{12\sqrt{10\,(1-\cmini)}}.
\ee
For $k=k_{\mathrm{min}}$ we get only one solution $\chii(k_{\mathrm{min}})=0$.

\subsubsection{Construction of the grids}
\label{G1construction}

We want to use the FFT algorithm in the computation of the $\F$-statistic
for all grid nodes, therefore we need such a grid that
(i) all grid points can be arranged along straight lines
parallel to the $\omega_0$-axis,
and (ii) the distance between neigbouring points along these lines is $k\pi$,
where $k=2^\ell$, $\ell=0,1,2,3$ [see Eq.\ \eqref{Domega0m}].
It is not difficult to construct an \emph{orthogonal} lattice
fulfilling these requirements.
Let the lattice be spanned by the vectors $(\mathbf{P}_0,\mathbf{P}_1)$.
The first basis vector $\mathbf{P}_0$ is parallel to the $\omega_0$-axis
and has components $\mathbf{P}_0:=(k\pi,0)$,
so its length is equal to the height of the fundamental parallelogram.
The second basis vector $\mathbf{P}_1$
is chosen to be parallel to the $\omega_1$-axis
and has length equal to the length of the parallelogram's base.
The $\omega_1$-component of the vector $\mathbf{P}_1$
can easily be obtained from Eqs.\ \eqref{204}.
One finds that $\mathbf{P}_1=(0,\delta\omega_1)$, where
\be
\label{212}
\delta\omega_1 =
\frac{\sqrt{15}\sqrt{48\,(1-\cmini)\,(1+60\chii^2)-k^2\,\pi^2}}
{1+60\chii^2}.
\ee

\begin{figure}
\begin{center}
\includegraphics[scale=0.60]{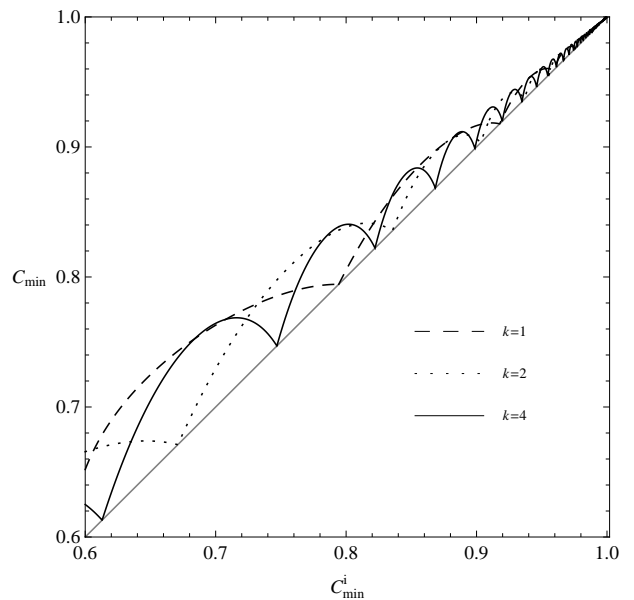}
\caption{\label{cmin_vs_cimin}
The relation, valid for the grids $G_1$,
between the initial minimum value $\cmini$
and the final minimum value $\cmin$ of the autocovariance.
Plots for three different values of the parameter $k$ are shown.
The diagonal grey line is the plot of the relation $\cmin=\cmini$.}
\end{center}
\end{figure}

In the construction of the fundamental parallelogram presented above
(in Appendix \ref{ParaEleCell})
we have ensured that the autocovariance between the center of the fundamental parallelogram
and any of its vertexes is equal to $\cmini$.
In the bank of templates spanned by the vectors $\mathbf{P}_0$
and $\mathbf{P}_1$ the minimum value [taken over all points in the $(\omega_0,\omega_1)$-plane]
of the maximum (taken over all grid points) autocovariance between any point in the
$(\omega_{0},\omega_{1})$-plane and all the grid points is not less than $\cmini$.
It turns out that usually it is \emph{greater} than $\cmini$.
Let us denote this minimum value of the autocovariance by $\cmin$,
then $\cmin\geqslant\cmini$.

To find the value of $\cmin$ we should first find the \emph{Voronoi cell} of the grid.\footnote{
With each point $\bxi_a$ of a lattice $\mathcal{L}$
we associate its \emph{Voronoi cell} $\mathrm{V}(\bxi_a)$
which consists of those points of plane that are at least as close to $\bxi_a$
as to any other point $\bxi_b\in\mathcal{L}$:
$\mathrm{V}(\bxi_a):=\{\bzeta: |\bzeta-\bxi_a| \le |\bzeta-\bxi_b| \:\text{for all}\: b \}$.}
To do this let us first choose two neighbouring grid nodes
and find all points in the $(\omega_0,\omega_1)$-plane
such that the autocovariance between the point and one of the chosen nodes
is equal to the autocovariance between the point and the second node.
All such points belong to the border between neighbouring Voronoi cells.
We repeat this construction for all nodes which are neighbours of the chosen node.
Points where borders cross each other are vertexes of the Voronoi cell
and at these vertexes the autocovariance takes its minimum value.
In Fig.\ \ref{cmin_vs_cimin} we have shown,
for different values of the parameter $k$,
the relation between the initial minimum value $\cmini$
and the final minimum value $\cmin$ of the autocovariance.

\begin{table*}
\caption{\label{tableG1}
Basis vectors $\mathbf{P}_0$ and $\mathbf{P}_1$
defining the orthogonal grids $G_{1,\ell}$ ($\ell=0,1,2,3$)
for $\cmin=0.75$ (which corresponds to $\cmine=0.7737$).
The covering thicknesses $\theta$ of the grids $G_{1,\ell}$
are shown in Table \ref{simula}.}
\begin{ruledtabular}
\begin{tabular}{cccrc}
$\ell$ & $\cmini$ & $\chii$ & $\mathbf{P}_0$ & $\mathbf{P}_1$ \\ \hline
0 & 0.68038480461 & 0.06911969777 & $( \pi,0)$ & $(0,9.4565868877)$ \\
1 & 0.71208037943 & 0.28027281768 & $(2\pi,0)$ & $(0,4.2593984076)$ \\
2 & 0.68138053992 & 0.57228732748 & $(4\pi,0)$ & $(0,2.3567814138)$ \\
3 & 0.69684651733 & 1.19594612580 & $(8\pi,0)$ & $(0,1.1211909237)$ \\
\end{tabular}
\end{ruledtabular}
\end{table*}

\subsection{Grids $G_2$}
\label{grid2&3}

Constructions of the grids $G_2$ were inspired
by the existence of the \emph{optimal hexagonal covering} by circles
of the 2-dimensional Cartesian space $\mathbb{R}^2$
and they can be treated as some deformations of this covering.
To employ the properties of the hexagonal covering we firstly
translate, by means of a linear transformation, the problem of covering
the $(\omega_0,\omega_1)$-plane by identical \emph{ellipses}
to the problem of covering the $(\omega'_0,\omega'_1)$-plane by \emph{unit circles}. 

\begin{figure}
\begin{center}
\includegraphics[scale=0.60]{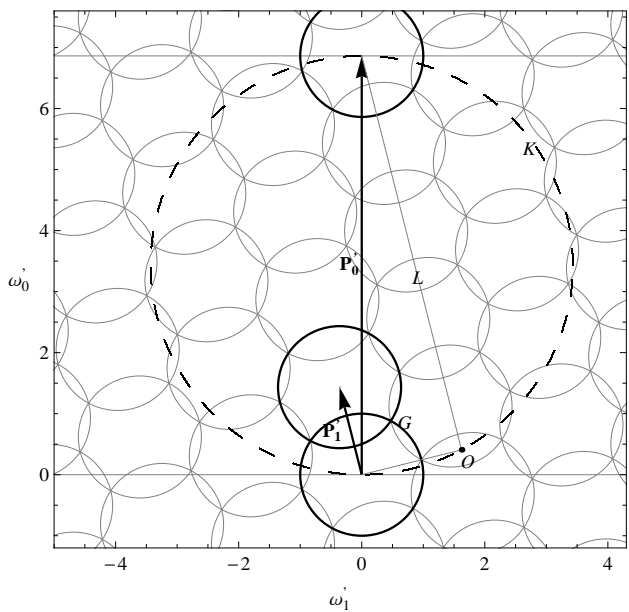}
\caption{\label{2ndmethod}
Construction of the grids $G_{2,\ell}$ (valid for $\ell\ge1$)
in the $(\omega'_0,\omega'_1)$-plane.}
\end{center}
\end{figure}

In the constructions of the grids described below
we were guided by the two following features of the hexagonal covering.
(i) The fundamental region of the hexagonal covering can be chosen to be
a regular hexagon inscribed into the circle; we demand that
deformed coverings have fundamental region in the form of
a polygon (usually nonregular hexagon) inscribed into the unit circle.
(ii) Let us denote by $(\mathbf{P}'_0,\mathbf{P}'_1)$ the basis
vectors of the lattice covering in the $(\omega'_0,\omega'_1)$-plane
(with the vector $\mathbf{P}'_0$ is parallel to the $\omega'_0$-axis);
we demand that the lattice points which lie along two neighbouring
straight lines parallel to the vector $\mathbf{P}'_1$
are shifted with respect to each other
by half of length of the vector $\mathbf{P}'_1$
(this feature can be seen in Fig.\ \ref{2ndmethod}).

We start from constructing a linear transformation
(described by a matrix $\mathsf{M}$)
which converts the ellipse of the autocovariance function
into the circle of unit radius.
The matrix $\mathsf{M}$ transforms a point with coordinates $(\omega_0,\omega_1)$
into the point with coordinates $(\omega'_0,\omega'_1)$:
\be
\label{tra21}
(\omega'_0,\omega'_1)^\mathsf{T}
= \mathsf{M}\cdot(\omega_0,\omega_1)^\mathsf{T}.
\ee
Equation of the ellipse of the autocovariance function
in the $(\omega_0,\omega_1)$-plane reads
[see Eq.\ \eqref{203}]
\be
\label{grids23a}
(\omega_0,\omega_1) \cdot \tilde{\Gamma} \cdot (\omega_0,\omega_1)^\mathsf{T}
= 1 - \cmin.
\ee
The linear transformation \eqref{tra21} converts the ellipse \eqref{grids23a}
into the circle of unit radius provided the matrix $\mathsf{M}$ fulfills the condition
\be
\label{grids23b}
\mathsf{M}^\mathsf{T} \cdot \mathsf{M}
= \frac{1}{1 - \cmin} \tilde{\Gamma}.
\ee
The Fisher matrix $\tilde{\Gamma}$ is symmetric and [what can easily be shown
by means of Eq.\ \eqref{034}] it is strictly positive definite, i.e.\
$(\omega_0,\omega_1)\cdot\tilde{\Gamma}\cdot(\omega_0,\omega_1)^\mathsf{T}>0$
for any $(\omega_0,\omega_1)\ne(0,0)$.
For such matrix $\tilde{\Gamma}$ the equation \eqref{grids23b}
can be interpreted as its Cholesky decomposition, which states that there
exists the unique upperdiagonal matrix $\mathsf{M}$ fulfilling Eq.\ \eqref{grids23b}.
In the rest of this subsection we will assume that the matrix $\mathsf{M}$
is the result of the Cholesky decomposition (so it is an upperdiagonal matrix).
Let us also note that the matrix $\mathsf{M}$ depends on the parameters $\chii$ and $\cmin$.

Let $\mathbf{P}_0$ be the vector in the $(\omega_0,\omega_1)$-plane
parallel to the $\omega_0$ axis and with length equal to $k\pi$,
so its $(\omega_0,\omega_1)$-components are $\mathbf{P}_0 = (k\pi,0)$.
After transformation to the $(\omega'_0,\omega'_1)$-plane this vector becomes $\mathbf{P}_0'$.
Let us denote its length by $ak\pi$,
thus its $(\omega'_0,\omega'_1)$-components are
(remember that the matrix $\mathsf{M}$ is upperdiagonal)
\be
\label{P0'}
\mathbf{P}'_0 = (ak\pi,0).
\ee

The construction of the second basis vector $\mathbf{P}_1'$
of the lattice in the $(\omega'_0,\omega'_1)$-plane
is described in details below. When this vector is found,
we take its image in the transformation inverse to that from Eq.\ \eqref{tra21},
it defines the second basis vector in the $(\omega_0,\omega_1)$-plane,
\be
\label{tra22}
\mathbf{P}_1 := \mathsf{M}^{-1}\mathbf{P}_1'.
\ee

For the lattices constructed below
the components of the basis vectors $\mathbf{P}_1$ depend on the parameter $\chii$,
therefore for different values of this parameter we get different lattices,
but, as we checked, all these lattices have (for the fixed $\ell$)
the same value of covering thickness.
Moreover, we can always choose such value of $\chii$
that the vectors $\mathbf{P}_0$ and $\mathbf{P}_1$ will be orthogonal.
In Table \ref{tableG2no} we give the components of the basis vectors $(\mathbf{P}_0,\mathbf{P}_1)$
for non-orthogonal grids $G_{2,\ell}$ (for $\ell=0,1,2,3$) defined
by choosing $\chii=0$. Table \ref{tableG2o} contains
the components of the vectors $(\mathbf{P}_0,\mathbf{P}_1)$
for orthogonal grids $G'_{2,\ell}$ (for $\ell=0,1,2,3$; we have added here primes
to the grid symbols to distinguish them from non-orthogonal grids of Table \ref{tableG2no})
together with the values of the parameter $\chii$ chosen to make the basis vectors orthogonal.

\begin{table}
\caption{\label{tableG2no}
Basis vectors $\mathbf{P}_0$ and $\mathbf{P}_1$
defining the non-orthogonal grids $G_{2,\ell}$ ($\ell=0,1,2,3$)
for $\cmin=0.75$ (which corresponds to $\cmine=0.7737$).
The grids are defined by choosing $\chii=0$.
The covering thicknesses $\theta$ of the grids $G_{2,\ell}$
are shown in Table \ref{simula}.}
\begin{ruledtabular}
\begin{tabular}{ccrc}
$\ell$ & $\chii$ & $\mathbf{P}_0$ & $\mathbf{P}_1$   \\ \hline
0 &  0 & $( \pi,0)$ & (1.57079632679490, 9.53468292515346) \\
1 &  0 & $(2\pi,0)$ & (1.91202057746303, 4.15260608129565) \\
2 &  0 & $(4\pi,0)$ & (2.65356333373361, 2.35185575858832) \\
3 &  0 & $(8\pi,0)$ & (2.92397504375649, 1.19974457106559) \\
\end{tabular}
\end{ruledtabular}
\end{table}

\begin{table}
\caption{\label{tableG2o}
Basis vectors $\mathbf{P}_0$ and $\mathbf{P}_1$
defining the orthogonal grids $G'_{2,\ell}$ ($\ell=0,1,2,3$)
for $\cmin=0.75$ (which corresponds to $\cmine=0.7737$).
The values of the parameter $\chii$ chosen to make the basis vectors orthogonal
are also given.
The covering thicknesses $\theta$ of the grids $G'_{2,\ell}$
are shown in Table \ref{simula}.}
\begin{ruledtabular}
\begin{tabular}{ccrc}
$\ell$ & $\chii$ & $\mathbf{P}_0$ & $\mathbf{P}_1$ \\ \hline
0 & 0.08237276158660 & $( \pi,0)$ & (0, 9.53468292515345) \\
1 & 0.23021935382642 & $(2\pi,0)$ & (0, 4.15260608129565) \\
2 & 0.56414244879677 & $(4\pi,0)$ & (0, 2.35185575858833) \\
3 & 1.21858231921793 & $(8\pi,0)$ & (0, 1.19974457106559) \\
\end{tabular}
\end{ruledtabular}
\end{table}

\subsubsection{Grids $G_{2,\ell}$ $($valid for $\ell\ge1)$}
\label{grid2}

To find the second basis vector $\mathbf{P}_1'$
in the $(\omega'_0,\omega'_1)$-plane,
we make the following construction, which is illustrated in Fig.\ \ref{2ndmethod}.
We plot a circle $K$ with radius equal to half of the length of the vector $\mathbf{P}'_0$.
The center of the circle coincides with the center of the line segment spanned by the vector $\mathbf{P}'_0$.
We inscribe a right-angled triangle $G$ into the circle $K$, its hypotenuse is along the diameter of the circle
(i.e.\ along the vector $\mathbf{P}'_0$).
The vertex $O$ at the right angle of the triangle $G$ has coordinates
\begin{align}
\label{210}
O = \bigg( \frac{1}{2}ak\pi-\sqrt{\frac{1}{4}(ak\pi)^{2}-{\omega'_1}^2},\omega'_1 \bigg).
\end{align}
The second basis vector $\mathbf{P}'_1$ is chosen to be parallel
to one of the sides of the triangle $G$, see Fig.\ \ref{2ndmethod}.
We demand that the ratio of the length $L$ of this side
and half of length of the vector $\mathbf{P}'_1$ is an odd positive integer,
$L/(|\mathbf{P}'_1|/2)=2n+1$ ($n=1,2,\ldots$).
For the fixed $n$ this requirement fixes the both coordinates of the point $O$
and thus it determines the vector $\mathbf{P}'_1$ uniquely.
Usually we obtain several possible values of $n$.
We choose this value which leads to lattice with the smallest covering thickness $\theta$.

\subsubsection{Grid $G_{2,0}$ $($valid for $\ell=0)$}
\label{grid3}

\begin{figure}
\begin{center}
\includegraphics[scale=0.6]{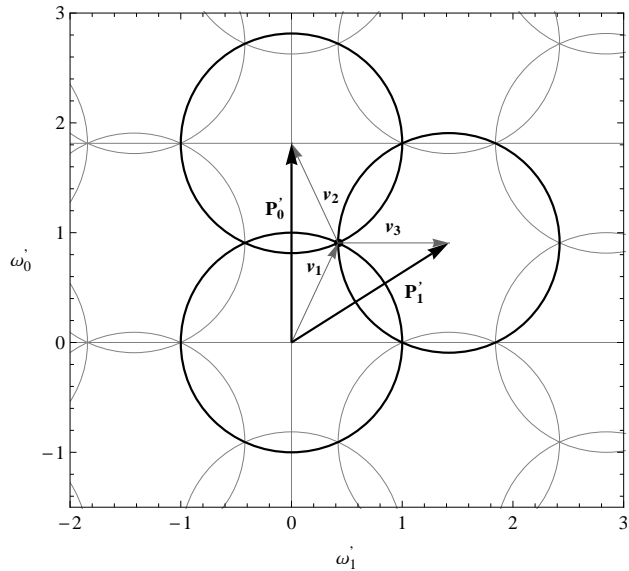}
\caption{\label{3circles}
Construction of the grid $G_{2,0}$ in the $(\omega'_0,\omega'_1)$-plane.}
\end{center}
\end{figure}

The construction of the grid $G_{2,0}$ in the $(\omega'_0,\omega'_1)$-plane
is illustrated in Fig.\ \ref{3circles}.
We start from constructing three circles of unit radii.
All these circles have to cross each other at the same point.
We build then the three vectors $\mathbf{v}_{1}$, $\mathbf{v}_{2}$, and $\mathbf{v}_{3}$
of unit length ($|\mathbf{v}_{1}|=|\mathbf{v}_{2}|=|\mathbf{v}_{3}|=1$).
The vector $\mathbf{v}_{1}$ begins at the centre of one of the circles
and ends at the point which is common to all three circles;
the vectors $\mathbf{v}_{2}$ and $\mathbf{v}_{3}$ are constructed in a similar way,
see Fig.\ \ref{3circles}.
The coordinates of the vectors $\mathbf{v}_{1}$, $\mathbf{v}_{2}$, and $\mathbf{v}_{3}$ are
\be
\label{218}
\mathbf{v}_{1}=(p,q), \quad
\mathbf{v}_{2}=(p,-q), \quad
\mathbf{v}_{3}=(0,1),
\ee
where $p$ and $q$ are positive numbers fulfilling the condition
\be
\label{213}
p^2+q^2=1.
\ee
Because (see Fig.\ \ref{3circles})
\be
\mathbf{P}'_0 = \mathbf{v}_{1}+\mathbf{v}_{2},
\ee
by virtue of Eq.\ \eqref{P0'} (taken for $k=1$) we get
\be
\label{216}
p = \frac{1}{2}a\pi.
\ee
Making use of Eqs.\ \eqref{213}, \eqref{216} and the equality
(see Fig.\ \ref{3circles})
\be
\label{217}
\mathbf{P}'_1 = \mathbf{v}_{1}+\mathbf{v}_{3},
\ee
one easily obtains the $(\omega'_0,\omega'_1)$-coordinates
of the basis vector $\mathbf{P}'_1$,
\be
\label{215}
\mathbf{P}'_1 = \Bigg(\frac{1}{2}\pi a, 1 + \sqrt{1-\Big(\frac{1}{2}\pi a\Big)^2}\Bigg).
\ee

\begin{acknowledgments}

The work presented in this paper was supported in part
by the Polish MNiSzW grant no.\ N N203 387237.
We would like to thank Andrzej Kr\'olak for helpful discussions.

\end{acknowledgments}

\end{document}